\documentclass[a4paper,11pt]{article}
\pdfoutput=1
\usepackage{jcappub}
\usepackage[T1]{fontenc}
\usepackage{amsmath,amssymb}
\usepackage{bm}
\usepackage{bbold}
\usepackage{paralist}

\newcommand{\rd}{\ensuremath{\mathrm{d}}}

\newcommand{\bfd}{\ensuremath{\bm{d}}}

\newcommand{\bft}{\ensuremath{\bm{t}}}

\newcommand{\bfalpha}{\ensuremath{\bm{\alpha}}}

\newcommand{\bfDelta}{\ensuremath{\bm{\Delta}}}

\newcommand{\bbR}{\ensuremath{\mathbb{R}}}
\newcommand{\bbeins}{\ensuremath{\mathbb{1}}}

\newcommand{\rmd}{\ensuremath{\mathrm{d}}}

\newcommand{\elL}{\ensuremath{\mathcal{L}}}

\title{On marginals and profiled posteriors for cosmological parameter
  estimation}

\author[a]{Martin Kerscher}
\author[a,b]{Jochen Weller}

\affiliation[a]{Ludwig--Maximilians Universt\"at M\"unchen,
  Universitäts-Sternwarte München, Fakultät für Physik,
  Scheinerstr.\,1, 81679 München, Germany
}
\affiliation[b]{Max Planck Institute for Extraterrestrial Physics,
  Giessenbachstr.\,1, 85748 Garching, Germany}
\emailAdd{martin.kerscher@lmu.de}

\abstract{
  With several examples and in an analysis of the Pantheon+ supernova
  sample we discuss the properties of the marginal posterior distribution 
  versus the profiled posterior distribution -- the profile likelihood in a
  Bayesian disguise.
  We investigate whether maximisation, as used for the profiling, or
  integration, as used for the marginalisation, is more appropriate.
  To report results we recommend the marginal posterior distribution.
}

\date{November 1, 2024}


\begin{document}
\maketitle

\section{Introduction}  

Cosmological models use several parameters like the matter content, a
dark energy component, or the Hubble parameter etc.  In a Bayesian
analysis, the posterior distribution of these parameters is determined
from astronomical observations (e.g.~\cite{trotta:sky}). In addition,
further nuisance parameters may enter the analysis. To focus on a
subset of the cosmological parameters, the marginal of the posterior
distribution is used to report the one-dimensional distribution
itself, the mean parameter values, or credible regions.
As an alternative, the profile likelihood has been used
\citep{herold:newconstraint, hamann:evidence, gonzales-morales:priors,
  ade:planckxvi}.  To focus on one parameter, one sets the other
parameters to the values maximising the likelihood.  The function of
the remaining parameter is called the profile likelihood. Together
with standard methods from classical statistics, this profile
likelihood can be used to determine the confidence region for a
parameter (see e.g.\ \cite{feldmann:unified}).
Profiling and marginalisation may give concordant and sometimes
differing results for the parameters. For the cosmic radiation density
this has been investigated by \cite{hamann:observational}, for an
early dark energy model see \cite{efstathiou:improved}, for an
entangled initial quantum state see \cite{adil:entanglement}, and for
the tensor-to-scalar ratio see \cite{campeti:new}.

It is common to attribute the differences between marginalisation and
profiling to a volume effect: If the maximum of the posterior is not
surrounded by a region with a large fraction of the posterior mass,
the marginalisation and profiling will give different results.
An example with a sum of two Gaussians illustrating the volume effect
was given by \cite{gomez-valent:fast}. She develops methods for a fast
detection of a volume effect.
Hadzhiyska et al.\,\cite{hadzhiyska:6parameter} use the Laplace
approximation centred on values of the profiled posterior to factorise
the full marginal posterior into the profiling part and their Laplace term.
Recently Raveria et al.\,\cite{raveri:understanding} use this
factorisation of \cite{hadzhiyska:6parameter} and are able to
distinguish the volume effect from the so called projection effect.
In some of our examples and for the supernova data analysis using the
flat\,$w$CDM and non-flat\,$\Lambda$CDM this Laplace approximation is
indistinguishable from the marginal posterior. However in addition we
study the limitations of this approximation in more complex correlated
settings.
Specifically for non-convex credibility regions the profiled posterior
and consequently the Laplace approximation may ignore a branch from
the full posterior. The maximum from the profiling may jump between
the branches, leading to discontinuities in the Laplace approximation
and to significant differences between the profiled and the marginalised
posterior.

In section\,\ref{sec:marginalised} we briefly review the basics of
parameter estimation. We construct a special posterior distribution
following a similar procedure as used for profiling the likelihood,
and call this the profiled posterior distribution -- the profile
likelihood in a Bayesian disguise (see also
\cite{hamann:observational,hadzhiyska:6parameter,colgain:mcmc}).
Following Hadzhiyska et al.\,\cite{hadzhiyska:6parameter} we give the
Laplace approximation to the marginalised posterior.  Then we
show that the profiled posterior is a marginal distribution of a
special Bayesian hierarchical model.
Both the profiled posterior and the marginal posterior distribution are
probability densities for the parameters under consideration.
Therefore, we can compare them directly.
Taking the perspective from classical statistics, that is hardly a
fair procedure, since we are missing the point, that in the classical
approach there is no distribution of the parameters.  This difference
between the classical and the Bayesian approach cannot be
reconciled. Here we stick to the Bayesian description, because this
reformulation allows us to compare the distributions obtained from
maximisation or from integration on an equal footing.
In section\,\ref{sec:example}, we discuss the properties of the
distributions for several examples. For a simple Gaussian model, the
marginal and profiled posterior agree. But we also consider a model
with a tuneable volume effect and a model with non-convex credible
regions where distinct differences show up.
In section\,\ref{sec:data} we compare the profiled and the marginal
posterior distributions of cosmological parameters, obtained from an
analysis of the supernova magnitude redshift relation using the
Pantheon+ supernova sample \cite{scolnic:pantheon}.  We use this to
highlight the differences between profiling and marginalisation.
In the CMB case as discussed in
\cite{hamann:observational,efstathiou:improved,adil:entanglement,campeti:new}
the differences between profiling and marginalisation are either small
or may be attributed to a volume effect. The shown credibility regions
are mainly convex.
In our analysis of the supernova sample we observe a pathology going
beyond the simple volume effect. There non-convex credibility regions
show up and the profiling and marginalisation lead to quantitatively and 
qualitatively different results.
In section\,\ref{sec:summary} we present arguments why we think that
the marginal posterior distributions should be the preferred way to
report results from a parameter estimation.

\section{Marginalisation and profiling}
\label{sec:marginalised}

We start with a classic setting for parameter estimation.
Consider a model with parameters
$(\theta_1,\ldots,\theta_d)= \Theta\in\bbR^d$. We constrain
these parameters with the observed data $\bfd$.  Bayes theorem gives
us the posterior distribution of the parameters
\begin{equation}
\label{eq:bayesupdate}
p(\Theta | \bfd) = \frac{\elL(\bfd | \Theta)\ p(\Theta)}{p(\bfd)}.
\end{equation}
The prior distribution $p(\Theta)$ is quantifying our prior knowledge 
about the parameters, the likelihood $\elL(\bfd | \Theta)$ is the 
probability of the data given a set of parameters, and $p(\bfd)$ is 
the normalisation, also called the evidence.
We are also interested in the distribution of only the $i$th parameter
$\theta_{i}$. The rest of the parameters is indicated by
$\Theta_{]i[}=(\theta_1,\ldots,\theta_{i-1},\theta_{i+1},
\ldots,\theta_d)\in\bbR^{d-1}$.  
Here we focus on \emph{one} parameter $\theta_i$, but analogously a
subset of parameters could be considered.
In the following, we will also write $\theta_{i}, \Theta_{]i[}$ together 
and identify this with the full set of the parameters $\Theta\in\bbR^d$.
The marginal posterior distribution is given by
\begin{equation}
  p(\theta_i | \bfd) = \int p(\theta_i, \Theta_{]i[} | \bfd)\ 
  \rmd \Theta_{]i[}.
  \label{eq:marginal} 
\end{equation}
In a Bayesian analysis, this marginal posterior can be used to derive a
credible region for $\theta_i$ or calculate its maximum or mean value.
The maximum likelihood estimate $\widetilde\Theta$ is defined by
\begin{equation}
  \widetilde\Theta=\text{arg max}_{\Theta}\ \elL(\bfd|\Theta).
\end{equation}
For an unimodal likelihood function, $\elL(\bfd|\Theta)$ the
$\widetilde\Theta$ is unique.  From the likelihood, we can also
determine the confidence regions using e.g.\ a likelihood ratio test
\cite{wassermann:all,feldmann:unified}.

\subsection{Profile likelihood}
The profile likelihood of the parameter $\theta_i$ is defined as the
function
\begin{equation}
  \label{eq:profilelike}
  \elL_p(\theta_i) = \max_{\Theta_{]i[}}\ \elL(\bfd | \theta_i, \Theta_{]i[}) .
\end{equation}
For each value of $\theta_i$ we determine the maximum of the
likelihood over the remaining parameters $\Theta_{]i[}$.
Correspondingly, the mapping
\begin{equation}
  \bft_{]i[}(\theta_i) = \text{arg max}_{\Theta_{]i[}}\
  \elL(\bfd | \theta_i, \Theta_{]i[})
  \label{eq:t}
\end{equation}
assigns to each parameter value $\theta_i\in\bbR$ the parameters
$\widehat\Theta_{]i[}\in\bbR^{d-1}$, where the likelihood is at its
maximum.
If the maximum is unique, the
$(\theta_i, \bft_{]i[}(\theta_i))\in\bbR^d$ defines the graph of
$\bft_{]i[}$ in the full parameter space. We call it the profiling
graph.
Now the profile likelihood can be written as
\begin{equation}
  \label{eq:likelihood-t}
  \elL_p(\theta_i) =  \elL(\bfd | \theta_i, \bft_{]i[}(\theta_i) ) .
\end{equation}
The conditioning on this graph does not necessarily lead to a new
likelihood.  But at least asymptotically, near the maximum, the
profile likelihood behaves like a likelihood and allows the
construction of approximate confidence sets \cite{murphy:onprofile}.
The profile likelihood $\elL_p(\theta_i)$ may be used to determine the
confidence interval around the maximum likelihood estimate of the
single parameter~$\widetilde\theta_i$.  No integration is used, but
with the maximisation procedure we follow the graph
$(\theta_i,\bft_{]i[}(\theta_i))$ in parameter space.

\subsection{Profiled posterior}
\label{sec:profiledpost}
To construct a Bayesian analogue of the profile likelihood, we use a
similar procedure as in Eqn.\,(\ref{eq:profilelike}), but now we are
profiling the posterior distribution $p(\Theta |\bfd)$. The function
\begin{equation}
  r(\theta_i)
  = \max_{\Theta_{]i[}}p(\theta_i, \Theta_{]i[} |\bfd) .
\end{equation}
is not a probability density, but since $r(\theta_i)\ge0$ we may
calculate the normalisation $c_i = \int r(\theta_i)\, \rmd \theta_i$
and define a probability density for $\theta_i$ by,
\begin{equation}
  p_p(\theta_i|\bfd) 
  = \frac{1}{c_i}\,  \max_{\Theta_{]i[}}p(\theta_i, \Theta_{]i[} |\bfd).
  \label{eq:profilepost}
\end{equation}
We call this $p_p(\theta_i|\bfd)$ the profiled posterior (see also
\cite{hamann:observational,hadzhiyska:6parameter,colgain:mcmc}).
For a flat prior $p(\Theta)=a$ and with Eqns.\,(\ref{eq:bayesupdate})
and (\ref{eq:likelihood-t}) we get,
\begin{equation}
  p_p(\theta_i|\bfd) 
  =  \frac{a}{c_i\ p(\bfd)}
  \max_{\Theta_{]i[}}  \elL(\bfd | \theta_i, \Theta_{]i[})
   = \frac{a}{c_i\ p(\bfd)} \elL(\bfd | \theta_i, \bft_{]i[}(\theta_i))
  \ \propto\  \elL_p(\theta_i).
  \label{eq:posterior-profile}
\end{equation}
Hence the profiled posterior $p_p$ is the Bayesian analogue of the
profile likelihood $\elL_p$. Throughout this article, we will use flat
priors because we want to emphasise the connection between the
profiled posterior distribution and the profile likelihood.
A flat prior $p(\Theta)=a$ on the full parameter space $\bbR^d$ is not
normalisable, but we can work around this. The argument above, showing
the proportionality of the profile likelihood and the profiled
posterior (\ref{eq:posterior-profile}), is still valid if
$\elL(\bfd|\cdot)$ has a finite support $A\subset\bbR^d$ and we choose
the uniform prior $p(\Theta)=a$ for $\Theta\in A$ and zero
otherwise. Then $1/a=\text{vol}(A)$ is the $d$--dimensional volume of
$A$.  This restriction of the prior to a set $A$ can also be done
approximately as long as $\elL(\bfd |\Theta)\ll1$ for
$\Theta\not\in A$.
The profiled posterior $p_p(\theta_i|\bfd)$ and the marginal posterior
$p(\theta_i|\bfd)$ are both probability densities for the parameter
$\theta_i$. This is the basis for the comparison done in 
section\,\ref{sec:example} and \ref{sec:data}.

\subsection{Laplace approximation}

In the vicinity of its maximum the posterior can be approximated by a
Gaussian which is called the Laplace approximation
\cite{kass:laplace,mackay:bayesian}.  Hadzhiyska et
al.\,\cite{hadzhiyska:6parameter} used a similar 
approximation to determine the posterior distribution by
\begin{equation}
  p(\theta_i, \Theta_{]i[} | \bfd) \approx  p_p(\theta_i| \bfd)\,
  \exp\left(-\tfrac{1}{2} \Delta^T {\cal F}(\theta_i)  \Delta \right) ,
\end{equation}
where  $\Delta=(\Theta_{]i[}-t_{]i[}(\theta_i))^T$ and 
\begin{equation}
  {\cal F}_{lm}(\theta_i) =
  \frac{1}{2} \frac{\partial^2}{\partial \phi_l \partial \phi_m} \log p(\theta_i, \Phi | \bfd) 
  \vert_{\Phi=t_{]i[}(\theta_i)}
\label{eq:hessian}
\end{equation}
with $\Phi=(\phi_1,\ldots,\phi_{d-1})$ for $l,m=1,\ldots,d-1$.
This is a Laplace approximation of the posterior with fixed $\theta_i$
around the value $t_{]i[}(\theta_i)$, as given by the profiling
procedure.  Hence the Gaussian integral over $\Theta_{]i[}$ in
Eqn.\,(\ref{eq:marginal}) can be performed, and is resulting in the
Laplace approximation $p_L(\theta_i|\bfd)$ of the marginal
$p(\theta_i|\bfd)$ as
\begin{align}
  p_L(\theta_i|\bfd)
  & =  C\  \frac{p_p(\theta_i|\bfd)}{\sqrt{\det {\cal F}(\theta_i)}} .
    \label{eq:laplace-marginal}
\end{align}
All terms independent of $\theta_i$ are merged into the normalisation
factor~$C$.
As illustrated by Hadzhiyska et al.\,\cite{hadzhiyska:6parameter} this
leads to a significant speed-up in calculations of the marginal posterior.
Even for complex posteriors this Laplace approximation can be a good
choice, as long as for each $\theta_i$ the posterior is close to a
Gaussian in the remaining parameters $\Theta_{]i[}$. We will show this
in the following examples and in the analysis of the supernova sample,
but we will also see that in certain situations the approximate
$p_L(\theta_i|\bfd)$ differs from $p(\theta_i|\bfd)$ significantly.

\subsection{The profiled posterior as the marginal of a
  Bayesian hierarchical model}
\label{sec:hierarchical}
Before we turn to the examples, we will show that the profiled
posterior, and consequently the profile likelihood, can be understood
as a marginal distribution of a special Bayesian hierarchical model.
In a hierarchical model, we use a further hyper-parameter $\phi$ to
model the dependencies between the parameters
$\Theta=(\theta_1,\ldots,\theta_d)$.  The extended model with
the hyper-parameter $\phi$ leads to the following Bayesian update
for the posterior distribution of $\Theta$ and $\phi$
\begin{equation}
  p(\Theta, \phi | \bfd)
  = \frac{\elL(\bfd | \Theta, \phi)\ p(\Theta,\phi)}{p(\bfd)}
  = \frac{\elL(\bfd | \Theta, \phi)\ p(\Theta |\phi) p_h(\phi) }{p(\bfd)}.
\end{equation}
The prior distribution $p(\Theta,\phi)$ factorises into the
conditional prior distribution $p(\Theta|\phi)$ of $\Theta$ depending
on $\phi$, and the hyper-prior $p_h(\phi)$.  Often the likelihood
$\elL(\bfd | \Theta, \phi)=\elL(\bfd | \Theta)$ is not directly
depending on the hyper-parameter $\phi$. Also, further levels are
possible, but for our application a simple one-level hierarchy is
enough.
The profiling procedure defines the mapping
$\bft_{]i[}(\theta_i)=\widehat\Theta_{]i[}$ (see Eqn.\,(\ref{eq:t}))
and allows us to determine the parameters $\widehat\Theta_{]i[}$
depending on the value of the parameter $\theta_i$.  We model this
strict dependency in the conditional distribution $p(\Theta |\phi)$
using Dirac~$\delta$ distributions
\begin{equation}
  \label{eq:hierarchical-prior}
  p(\Theta |\phi)
  = \delta(\phi-\theta_i)\ \delta(\bft_{]i[}(\theta_i)-\Theta_{]i[}),
\end{equation}
with $\Theta=(\theta_i,\Theta_{]i[})$.
Given $\phi$, this definition specifies a probability distribution for
$\Theta$, since $p(\Theta |\phi)\ge0$ for any $\Theta$, and also
$\int p(\Theta |\phi) \rd\Theta = 1$.
As before, we are interested in the distribution of $\theta_i$, hence
we integrate over $\Theta_{]i[}$ and also the hyper-parameter $\phi$ to
obtain the marginal posterior $p_h(\theta_i|\bfd)$ for this
hierarchical model.
\begin{align}
  p_h(\theta_i | \bfd)
  &= \int_{\bbR^{d-1}} \int_{\bbR}\ p(\theta_i,\Theta_{]i[}, \phi | \bfd)\
    \rmd\phi\,\rmd\Theta_{]i[}\nonumber\\
  &= \int_{\bbR^{d-1}} \int_{\bbR}\ \frac{1}{p(\bfd)}\, 
    \elL(\bfd | \theta_i,\Theta_{]i[})\ \delta(\phi-\theta_i)\
    \delta(\bft_{]i[}(\theta_i)-\Theta_{]i[})\ p_h(\phi)\ \rmd\phi\,\rmd\Theta_{]i[}\nonumber\\
  & = \frac{1}{p(\bfd)}\ \elL(\bfd | \theta_i,\bft_{]i[}(\theta_i))\ p(\theta_i)
    = \frac{p(\theta_i)}{p(\bfd)}\ \elL_p(\theta_i).
    \label{eq:hierarchical}
\end{align}
Here we assume that the likelihood has no explicit dependence on the
hyper-parameter $\phi$ and that the prior distribution for the
parameter $\theta_i$ is equal to the hyper-prior
$p_h(\theta_i)=p(\theta_i)$.  If we assume a flat prior for the
parameters, we see that the marginal of this hierarchical model is
equal to the profiled posterior of Eqn.\,(\ref{eq:posterior-profile}):
\begin{equation}
  p_h(\theta_i | \bfd) = p_p(\theta_i|\bfd).
\end{equation}
Hence the marginal posterior of this hierarchical model is equal
to the profiled posterior distribution, the Bayesian analogue of the
profile likelihood.
Equation\,(\ref{eq:hierarchical-prior}) shows that for the profiled
posterior distribution a rather restrictive prior is used: the
parameter $\theta_i$ is acting as the hyper-parameter, and we have a
rigid dependence of the parameters $\Theta_{]i[}$ on the parameter of
interest $\theta_i$ through the mapping $\bft_{]i[}()$ as given in
equation\,(\ref{eq:t}).

\section{Example Distributions}
\label{sec:example}

We illustrate the effect of profiling the posterior with several
two-dimensional examples and compare the profiled posterior to the
marginal posterior distribution.

\subsection{A Gaussian Distribution}

\begin{figure}
  \begin{minipage}{0.4\textwidth}
    \includegraphics[width=\textwidth]{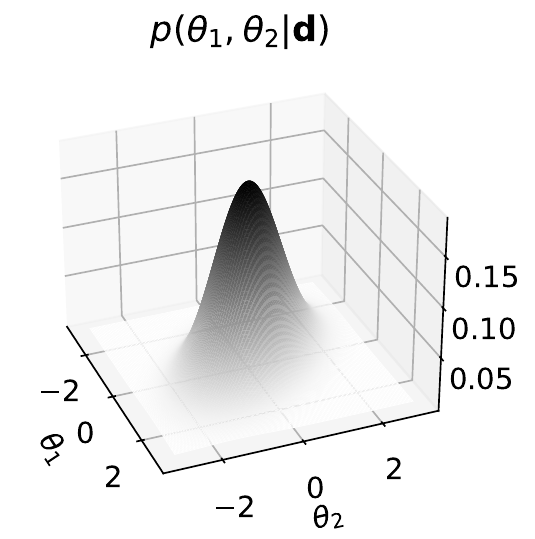}\\
    \includegraphics[width=\textwidth]{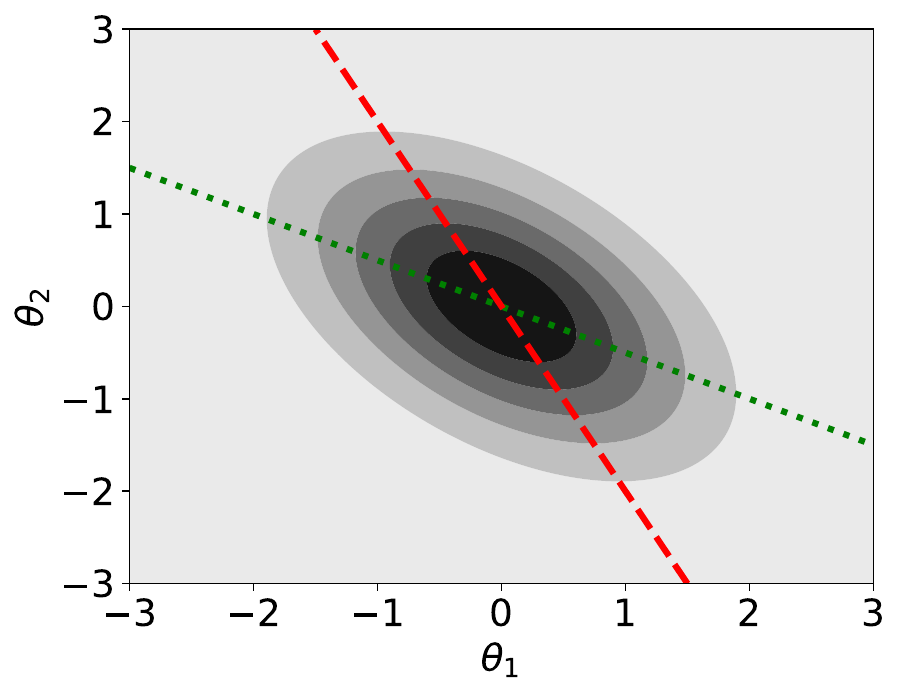}
  \end{minipage}
  \hfill
  \begin{minipage}{0.55\textwidth}
    \includegraphics[width=0.9\textwidth]{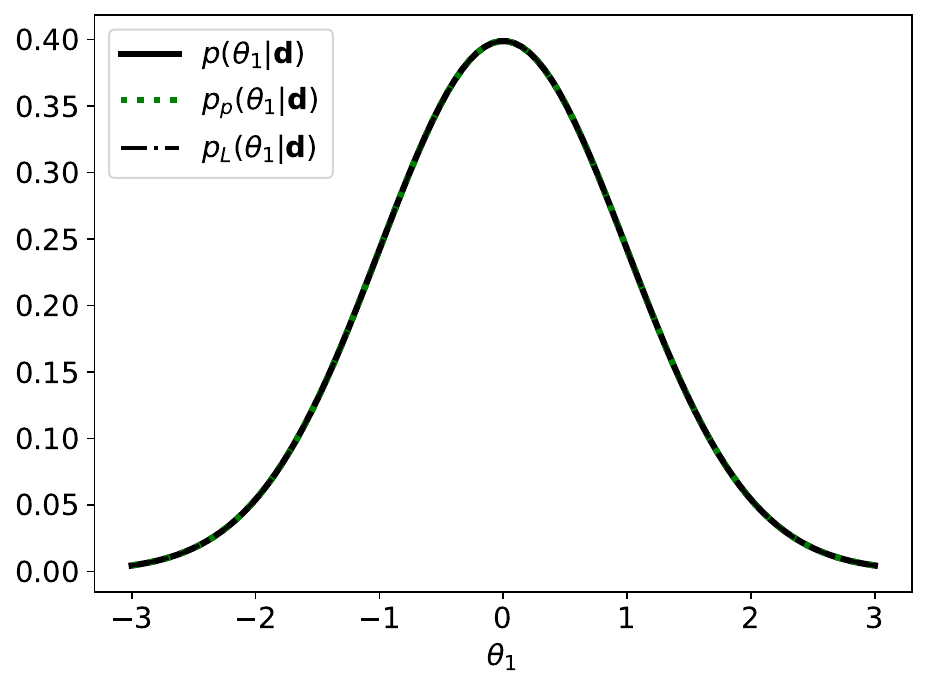}\\
    \includegraphics[width=0.9\textwidth]{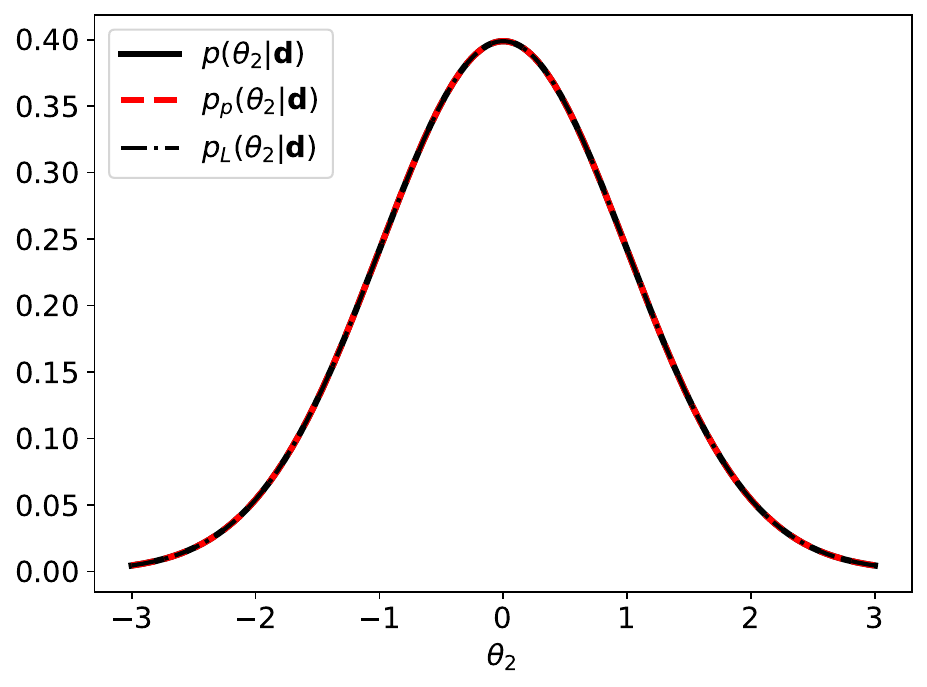}
  \end{minipage}
  \caption{On the left, the two-dimensional Gaussian posterior
    $p(\theta_1,\theta_2|\bfd)$ from Eqn.\,(\ref{eq:simple}) is shown
    with correlation coefficient $\rho=-0.5$. In the contour plot, we
    also show the profiling graphs $(\theta_1, \bft_{]1[}(\theta_1))$
    (green dotted) and $(\bft_{]2[}(\theta_2),\theta_2)$ (red
    dashed). On the right on top, we compare the marginal
    $p(\theta_1|\bfd)$ to the profiled posterior
    $p_p(\theta_1|\bfd)$. On the right, at the bottom, we compare the
    marginal $p(\theta_2|\bfd)$ with the profiled posterior
    $p_p(\theta_2|\bfd)$. In both cases, the marginal and the profiled
    posterior distribution overlap. Also the Laplace approximations
    $p_L(\theta_i|\bfd)$ are indistinguishable from the corresponding
    marginal distributions.}
  \label{fig:simple_example}
\end{figure}

As a first simple case, we consider a posterior distribution of
the parameters $\theta_1$ and $\theta_2$ given by a two-dimensional
Gaussian distribution
\begin{equation}
  p(\theta_1,\theta_2|\bfd ) = \frac{1}{2\pi\sqrt{(1-\rho^2)}}
  \exp\left(- \tfrac{1}{2 (1-\rho^2)}
    \left(\theta_1^2 -2\rho\theta_1\theta_2 +\theta_2^2\right) \right),
  \label{eq:simple}
\end{equation}
with unit variance and correlation coefficient $\rho$.  Envision this
as a posterior built from a Gaussian likelihood together with a flat
prior. Hence including $\bfd$ in the equation above is meant as a
reminder that the posterior is constructed from a model for the data
using a likelihood.

We calculate the marginal posterior distributions $p(\theta_1|\bfd)$,
$p(\theta_2|\bfd)$ according to Eqn.\,(\ref{eq:marginal}) and the
profiled posteriors $p_p(\theta_1|\bfd)$, $p_p(\theta_2|\bfd)$ from
Eqn.\,(\ref{eq:profilepost}). The marginal distributions of a
two-dimensional Gaussian are one-dimensional Gaussians, and as can be
seen from figure\,\ref{fig:simple_example}, the profiled posterior
distributions are identical to these Gaussian marginal distributions.

To calculate the Laplace approximation we determine the $\cal F$ from
Eqn.\,(\ref{eq:hessian}) with finite differences followed by a
Richardson extrapolation as provided by
numdifftools~\cite{brodtkorb:numdifftools}.  For this Gaussian example
a numerical approximation would not have been necessary, but we will
follow this numerical scheme also for the following more complicated
examples. Clearly, the Laplace approximation
Eqn.\,(\ref{eq:laplace-marginal}) is exact for a Gaussian as can be
also seen from figure~\ref{fig:simple_example}.

The profiled posterior distribution is the distribution of one
parameter, given that the other parameter is following this profiling
graph (see Eqn.\,(\ref{eq:hierarchical})). 
%
Both profiling graphs are straight lines.
The $\bft_{]1[}$ maps each $\theta_1$ to a
$\theta_2=\bft_{]1[}(\theta_1)$, and the profiling graph consists out
of these points $(\theta_1,\theta_2=\bft_{]1[}(\theta_1))$.
Similarly, for the profiling graph $(\bft_{]2[}(\theta_2),\theta_2)$,
each $\theta_2$ is mapped to a $\theta_1=\bft_{]2[}(\theta_2)$.

\subsection{A model for the volume effect}
\label{sec:veff-model}

\begin{figure}
  \begin{minipage}{0.4\textwidth}
    \includegraphics[width=\textwidth]{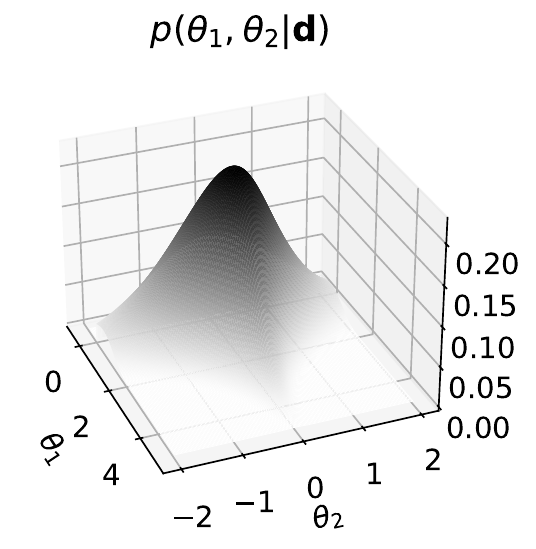}
    \includegraphics[width=\textwidth]{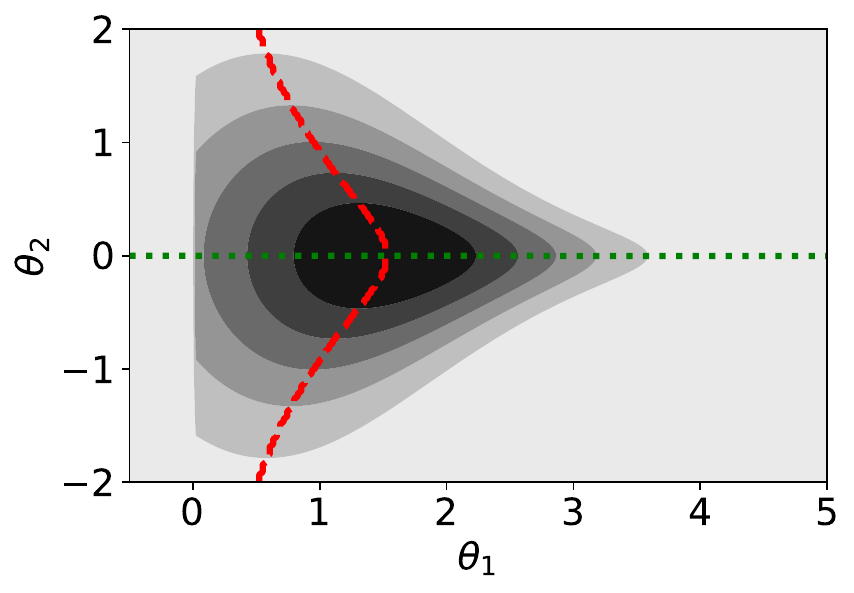}
  \end{minipage}
  \hfill
  \begin{minipage}{0.55\textwidth}
    \includegraphics[width=0.9\textwidth]{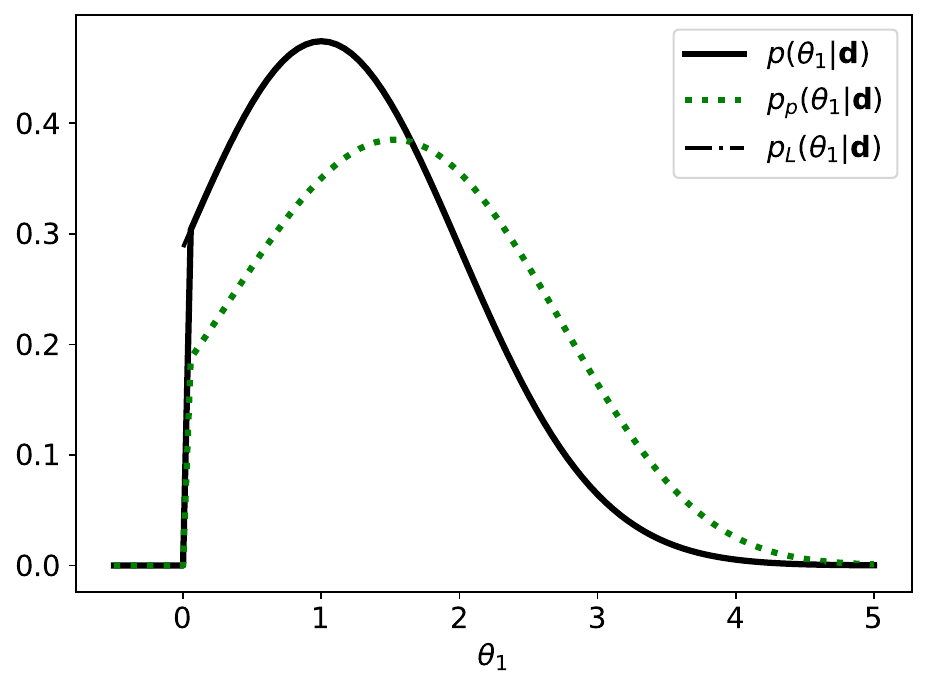}\\
    \includegraphics[width=0.9\textwidth]{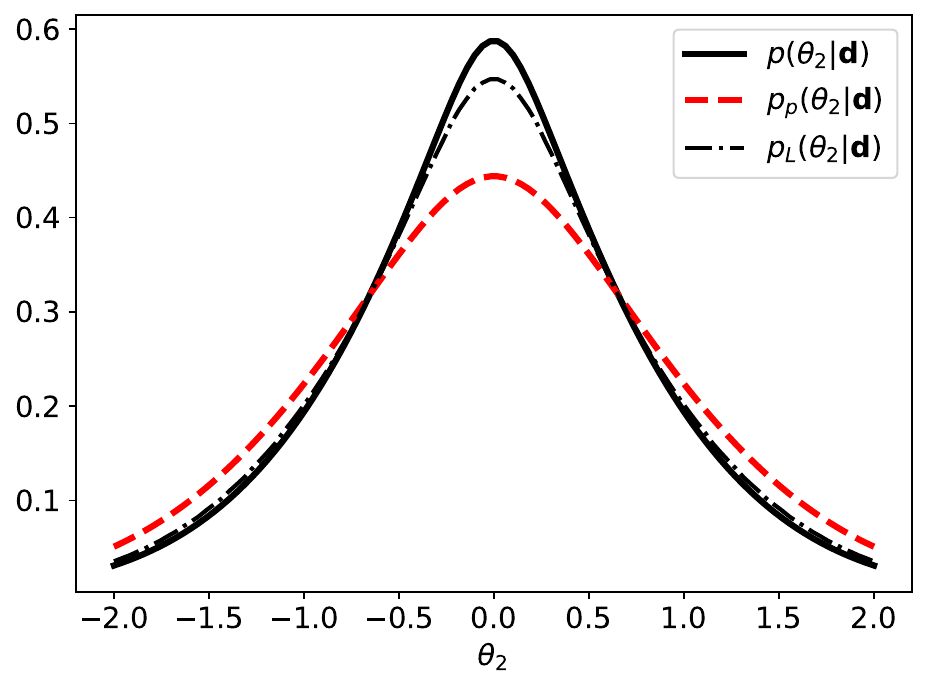}
  \end{minipage}
  \caption{On the left, the two-dimensional posterior
    $p(\theta_1,\theta_2|\bfd)$ is shown for the example illustrating
    the volume effect with $b=0.2$ (see Eqn.\,(\ref{eq:q2_b})). In the
    contour plot, we also show the profiling graphs
    $(\theta_1, \bft_{]1[}(\theta_1))$ (green dotted) and
    $(\bft_{]2[}(\theta_2),\theta_2)$ (red dashed). On the right on
    top, we compare the marginal $p(\theta_1|\bfd)$ to the profiled
    posterior $p_p(\theta_1|\bfd)$. On the right, at the bottom, we
    compare the marginal $p(\theta_2|\bfd)$ with the profiled
    posterior $p_p(\theta_2|\bfd)$. The corresponding Laplace approximations
    $p_L(\theta_i|\bfd)$ are shown as dashed-dotted lines.}
  \label{fig:veff_example}
\end{figure}

The following construction mimics the situation in cosmological
parameter estimation if our cosmological model~B is an extension of
the reference model~A but with two additional parameters.  Model~A is
determined by the parameters, $\bfalpha$ which we will not consider
further. In addition, to these parameters, model~B has two further
parameters $\theta_1\ge0$ and $\theta_2$.  For $\theta_1=0$ model~B
turns into model~A.
The volume effect in such a nested model can be described in the
following way:
When the parameter $\theta_1$ approaches zero the model~B reduces to
model~A, and in this limit the other parameter $\theta_2$ of model~B 
is less constrained. This leads to an increased volume in the
parameter space for
model~A, which could drive the posterior towards low values of
$\theta_1$ upon marginalisation.
See also Herold \& Ferreira \cite{herold:resolving} in their
presentation of the volume effect for the early dark energy model.

We construct the following model showing the traits of
the volume effect as described above.
The distribution of the parameter $\theta_1\ge0$ is assumed to be a
truncated Gaussian with mean $a$ and variance $\sigma_1^2$:
\begin{align}
  q_1(\theta_1)
  & = C \exp\left(- \frac{(\theta_1-a)^2}{2 \sigma_1^2}\right)\
    \text{for } \theta_1>0
\end{align}
and $q_1(\theta_1)=0$ for $\theta_1\le0$. The normalisation is
$1/C=\int_0^\infty\exp\left(- \frac{(\theta_1-a)^2}{2
    \sigma_1^2}\right)\rmd \theta_1$.
The distribution of the parameter $\theta_2$ is again a Gaussian with
variance $\sigma_2^2$.  For large $\theta_1\gg0$ we assume a constant
$\sigma_2\approx v$ corresponding to a finite resolution of the
measurement or observations. For small $\theta_1$ the distribution
of $\theta_2$ should broaden. We model this behaviour with
\begin{align}
  q_2(\theta_2|\theta_1) & = \frac{1}{ \sigma_2 \sqrt{2\pi}}
   \exp\left(- \frac{\theta_2^2}{2 \sigma_2^2}\right)\quad
   \text{and}\quad \sigma_2 = \exp(-\,b\, \theta_1^2) + v.
   \label{eq:q2_b}                        
\end{align}
Now the likelihood can be written as
\begin{equation}
  \elL( \bfd |\theta_1,\theta_2) = q_2(\theta_2|\theta_1)\ q_1(\theta_1).
  \label{eq:veff}
\end{equation}
Comparing Eqn.\,(\ref{eq:veff}) to  Eqn.\,(\ref{eq:berger3}) from the 
next section, we realise that this example is closely related to the 
ridge example of Berger et al.\,\cite{berger:integrated}. For another model 
illustrating the volume effect, see \cite{gomez-valent:fast}.

To detail our example further, we assume $a=1$, $\sigma_1=1$, and
$v=0.1$, $b=0.2$.  We want to compare a well-defined profiled
posterior to the profile likelihood, therefore we have to use a flat
prior restricted to a finite parameter domain.  In the following, we
choose $\theta_1\in A_1=[0,8]$ and $\theta_2\in A_2=[-5,5]$.  The
posterior is then
$ p(\theta_1,\theta_2| \bfd)\propto \elL( \bfd |\theta_1,\theta_2)
\bbeins_{A_1}(\theta_1)\bbeins_{A_2}(\theta_2)$ (with $\bbeins_A(x)=1$
if $x\in A$ and zero otherwise).
For each $\theta_1$, we determine the function
$r(\theta_1)=\max_{\theta_2} p(\theta_1,\theta_2| \bfd)$ by
numerical maximisation. Then we determine the normalisation
$c_1=\int_{A_1} r(\theta_1) \rmd \theta_1$ by a numerical integration
and finally obtain
\begin{equation}
  p_p(\theta_1| \bfd) =
  \tfrac{1}{c_1}\max_{\theta_2} p(\theta_1,\theta_2| \bfd).
\end{equation}
We proceed similarly for $p_p(\theta_2|\bfd)$.  With a numerical
integration of Eqn.\,(\ref{eq:veff}) we determine the marginal
posterior distributions according to Eqn.\,(\ref{eq:marginal}).

In figure\,\ref{fig:veff_example} the bulge for small $\theta_1$ is
clearly visible in the two-dimensional distribution
$p(\theta_1,\theta_2| \bfd)$. Due to the mirror symmetry the profiling
graph, $(\theta_1, \bft_{]1[}(\theta_1)=0)$ is a straight line,
whereas the profiling graph $(\bft_{]2[}(\theta_2),\theta_2)$ is
showing a dent.
The profiled posterior and marginal posterior distributions for
$\theta_2$ are symmetric around the same maximum, but the marginal
posterior is more strongly peaked.
For $\theta_1$ the profiled posterior is showing a significant shift
towards large values. The maximum of the marginal posterior is at
$\theta_1=1$ but the maximum of the profiled posterior is at the
larger value $\theta_1=1.67$.  Also, the marginal posterior
distribution has significantly more weight for smaller values of
$\theta_1$ than the profiled posterior distribution.  This behaviour is
called the "volume effect".
For a fixed $\theta_1$ the likelihood\,(\ref{eq:veff}) is a Gaussian
in $\theta_2$ and consequently the $p_L(\theta_1|\bfd)$ is giving the
exact value for the marginal $p(\theta_1|\bfd)$. For $\theta_2$ the
Laplace approximation $p_L(\theta_2|\bfd)$ is slightly broader than
the marginal.

\begin{figure}
  \begin{minipage}{0.49\textwidth}
    \includegraphics[width=0.49\textwidth]{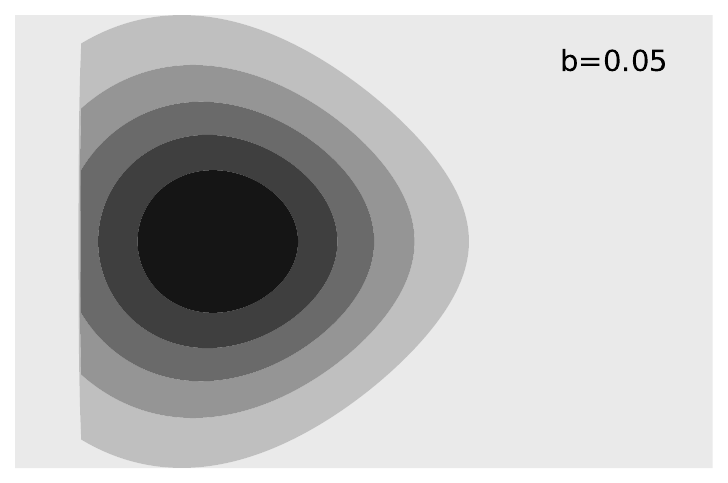}
    \includegraphics[width=0.49\textwidth]{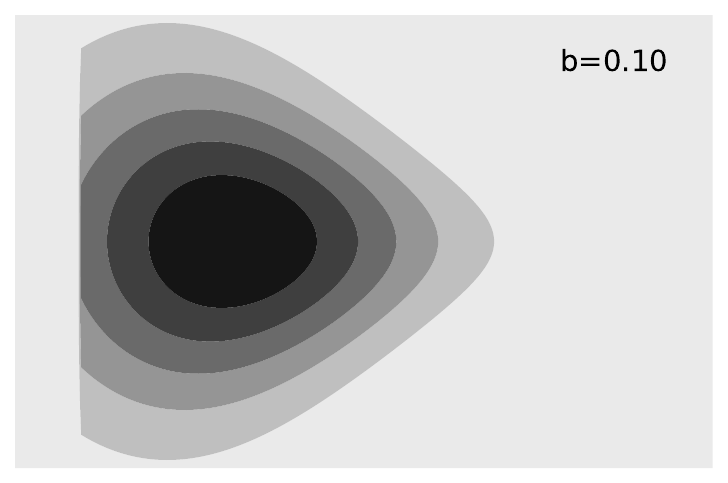}\\
    \includegraphics[width=0.49\textwidth]{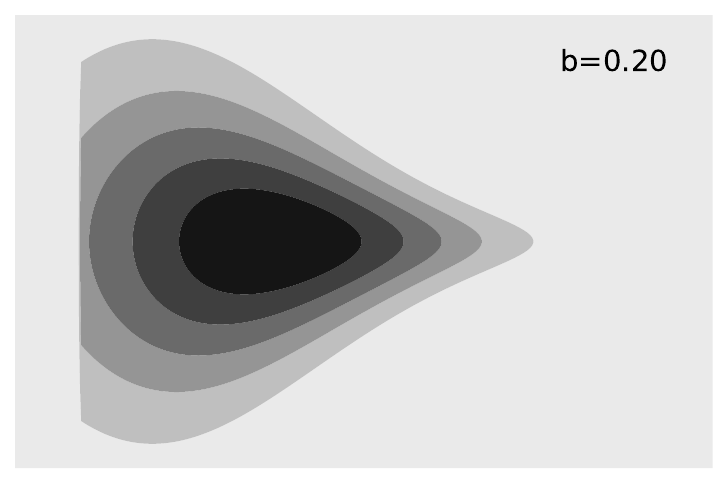}
    \includegraphics[width=0.49\textwidth]{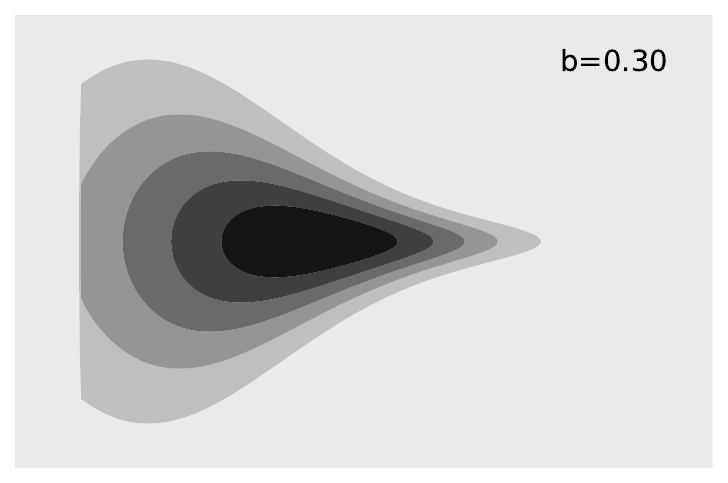}\\
  \end{minipage}
  \hfill
  \begin{minipage}{0.49\textwidth}
    \includegraphics[width=\textwidth]{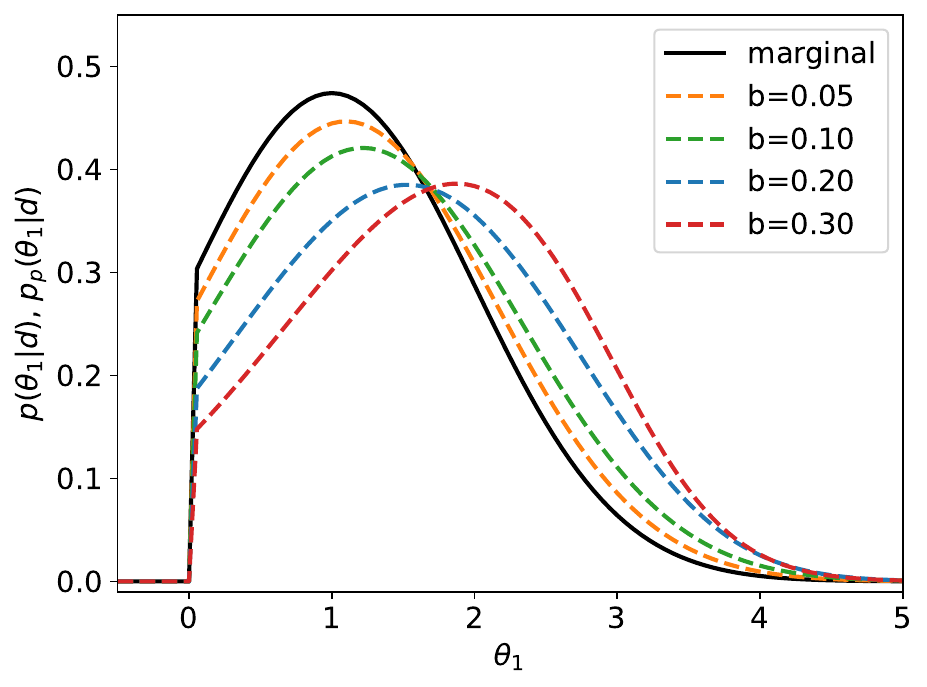}
  \end{minipage}
  \caption{ The contour plots on the left show the
    $p(\theta_1,\theta_2| \bfd)$ for several values for $b$ (compare
    also with figure\,\ref{fig:veff_example}).  In the right plot the
    marginal posterior $p(\theta_1|\bfd)$ (solid black line) and
    the profiled posterior distributions $p_p(\theta_1|\bfd)$ (dashed
    lines) are shown for a varying factor $b$ according to
    Eqn.(\ref{eq:q2_b}). The marginal is invariant under changes of $b$.}
  \label{fig:veff_dependence_b}
\end{figure}

With the factor $b$ in Eqn.\,(\ref{eq:q2_b}) we specify how strongly
the variance of the parameter $\theta_2$ depends on the parameter
$\theta_1$.  In this way, we obtain a model with a tuneable volume
effect.  Figure\,\ref{fig:veff_dependence_b} shows the marginal
$p(\theta_2|\bfd)$ and the profiled posterior distribution
$p_p(\theta_2|\bfd)$ for a series of values for $b$. The marginal
posterior distribution is independent of the value of $b$, whereas the
profiled posterior distribution is strongly depending on the value of
$b$.  For small $b$ the profiled posterior is approaching the marginal
posterior distribution, but for larger $b$ we generate a strong
"volume effect" and the profiled posterior distribution is shifted
towards larger values for $\theta_1$.
Similarly, we considered a variable $v$ in Eqn.,(\ref{eq:q2_b}) for a
fixed $b$. In this way, we investigate how the resolution of the
measurements influences the distribution. Essentially, the same
picture as in figure\,\ref{fig:veff_dependence_b} emerges: the marginal
posterior distribution $p(\theta_2|\bfd)$ does not depend on the value
of $v$, while the profiled posterior distribution $p_p(\theta_2|\bfd)$
is strongly dependent on $v$. For large $v$ the profiled posterior
approaches the marginal posterior distribution
 
In this model, we may tune the volume effect or the resolution.  In a
typical situation of parameter estimation, both are fixed by the
experimental setup. Hence, it seems advisable to use the marginal
posterior distribution if one wants to get results independent of the
resolution or the details of a volume effect.

\subsection{A model with a ridge}
\label{sec:ridge}

Berger et al.\,\cite{berger:integrated} discuss with several examples
why an integrated (marginalised) likelihood should be preferred over a
profile likelihood. Here we give a slightly simplified
version\footnote{To make the connection with Berger et
  al.\,\cite{berger:integrated} use $n=1$ and $\theta_2=\lambda$ in
  their Eqn.\,(11),\,(12), and~(13).}  of their example~3. Our model
from the last section is closely related to this ridge model. We
present the construction of this model in some detail, since it allows
us to show that the resulting likelihood is not an ad hoc structure.
Consider two measurements modelled with two independent random
variables $X$ and $Y$, where $X\sim N(\theta_1,1)$ follows a normal
distribution with mean~$\theta_1$ and unit variance, and
$Y\sim N(\theta_2,\exp(-\theta_1^2))$ with mean $\theta_2$ and
variance that depends on $\theta_1$ exponentially. The likelihood can
be obtained by multiplying the two Gaussians resulting in
\begin{align}
  \elL( \bfd |\theta_1,\theta_2)
  & = \frac{1}{\sqrt{2\pi}} \exp\left(- \tfrac{1}{2}(x-\theta_1)^2\right)\ 
    \frac{1}{\sqrt{2\pi\exp(-\theta_1^2)}} 
    \exp\left(- \frac{(y-\theta_2)^2}{2\exp(-\theta_1^2)}\right) \nonumber \\
  & = \frac{1}{2\pi}
  \exp\left(- \tfrac{1}{2}\left(x^2-2x\theta_1\right) -
    \frac{(y-\theta_2)^2}{2\exp(-\theta_1^2)} \right).
  \label{eq:berger3}
\end{align}
The data $\bfd=(x,y)$ consists out of the two measured 
values $x$ and $y$.
The conditional maximum likelihood estimate for $\theta_2$ is
$\widetilde\theta_2=y$ for each $\theta_1$, therefore, the profile
likelihood for $\theta_1$ is
$\elL_p(\theta_1) \propto \exp\left(x\theta_1\right)$ (see Berger et
al.\,\cite{berger:integrated}).  Depending on the sign of the
measured value $x$, this profile likelihood $\elL_p(\theta_1)$ is
exponentially divergent for $\theta_1>0$ if $x>0$ (or for $\theta_1<0$ if
$x<0$).  However, a marginalised likelihood reads
$p(\theta_1|\bfd)\propto \exp(-\tfrac{1}{2}(x-\theta_1)^2)$ providing
some localisation in parameter space.

To concretise, we assume that we observed $x=1$ and $y=0$, hence
$\bfd=(1,0)$. As above we use a flat prior restricted to a finite
parameter domain.  In the following, we choose $\theta_1\in A_1=[-1,1]$
and $\theta_2\in A_2=[-1,4]$.  The posterior is then
$ p(\theta_1,\theta_2| \bfd)\propto \elL( \bfd |\theta_1,\theta_2)
\bbeins_{A_1}(\theta_1)\bbeins_{A_2}(\theta_2)$ (with $\bbeins_A(x)=1$
if $x\in A$ and zero otherwise).
We calculate the profiled and marginal posterior distributions as
explained in section\,\ref{sec:veff-model}.
\begin{figure}
  \begin{minipage}{0.4\textwidth}
    \includegraphics[width=\textwidth]{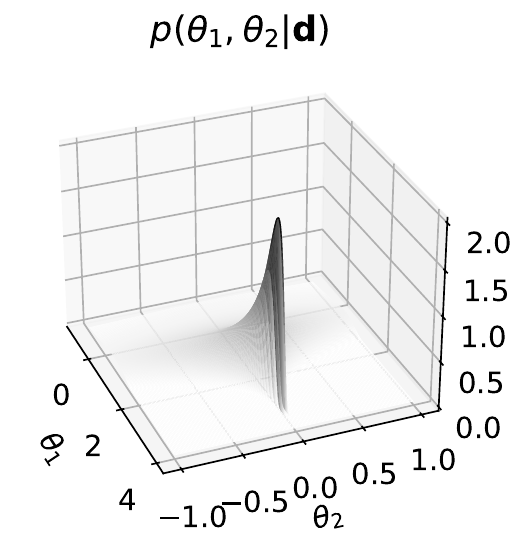}
    \includegraphics[width=\textwidth]{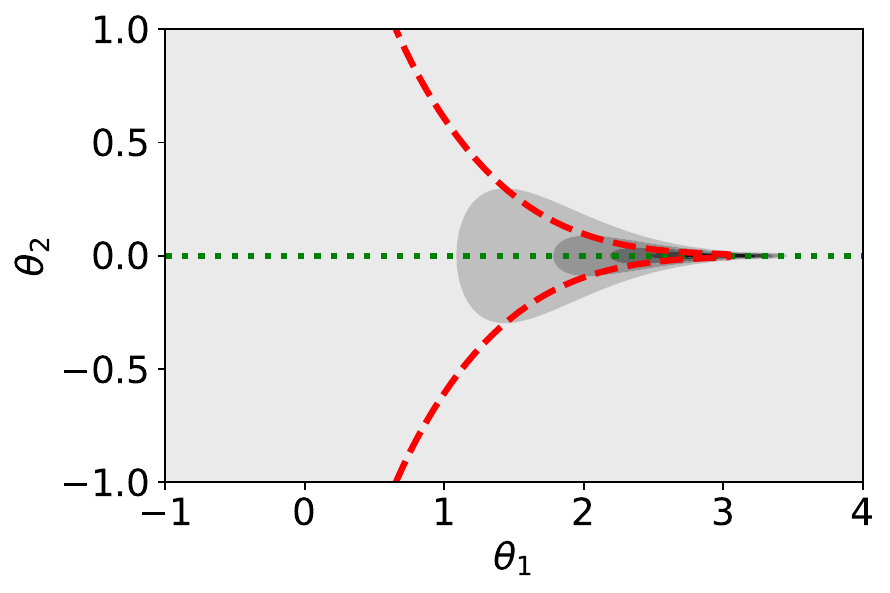}
  \end{minipage}
  \hfill
  \begin{minipage}{0.55\textwidth}
    \includegraphics[width=0.9\textwidth]{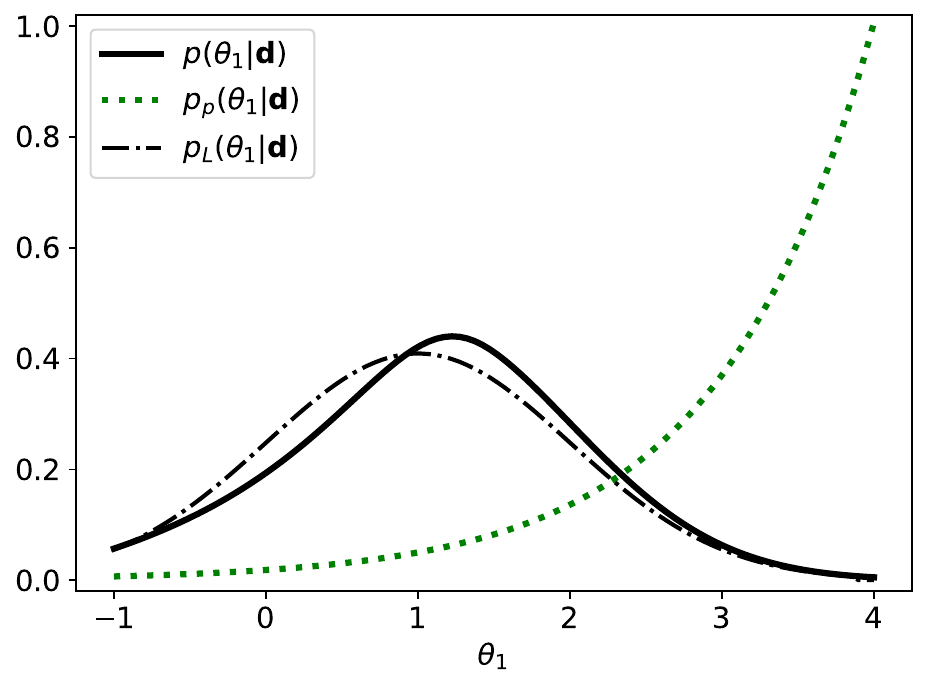}\\
    \includegraphics[width=0.9\textwidth]{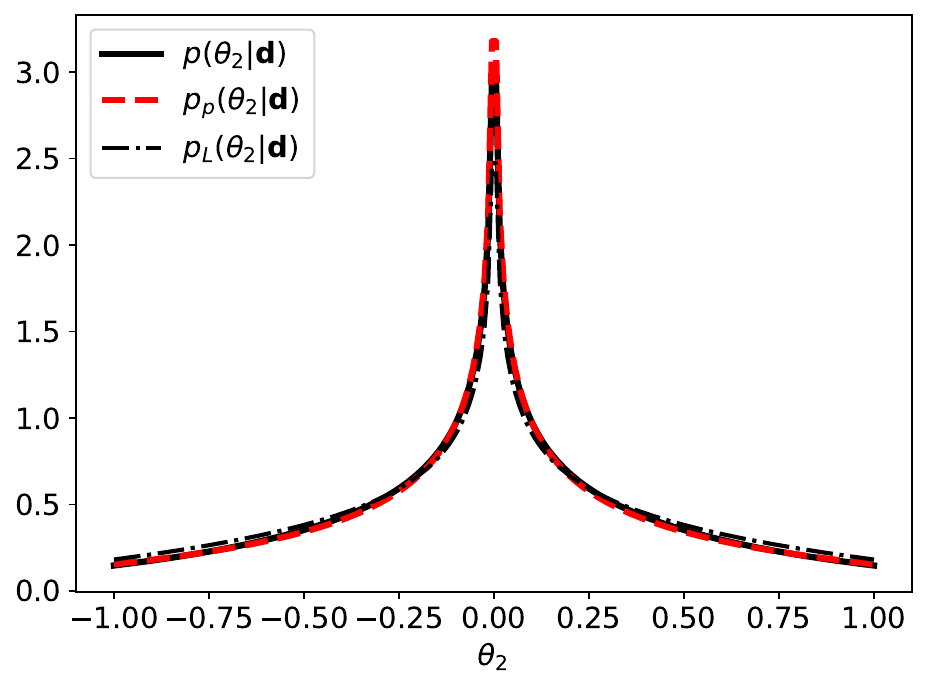}
  \end{minipage}
  \caption{On the left, the two-dimensional posterior
    $p(\theta_1,\theta_2|\bfd)$ is shown for the ridge example. In the
    contour plot, we also show the profiling graphs
    $(\theta_1, \bft_{]1[}(\theta_1))$ (green dotted) and
    $(\bft_{]2[}(\theta_2),\theta_2)$ (red dashed). On the right on
    top, we compare the marginal $p(\theta_1|\bfd)$ to the profiled
    posterior $p_p(\theta_1|\bfd)$. On the right, at the bottom, we
    compare the marginal $p(\theta_2|\bfd)$ with the profiled
    posterior $p_p(\theta_2|\bfd)$. The corresponding Laplace approximations
    $p_L(\theta_i|\bfd)$ are shown as dashed-dotted lines. }
  \label{fig:berger_example}
\end{figure}
In figure\,\ref{fig:berger_example} the ridge is clearly visible in
the distribution of the parameters. Due to the mirror symmetry
the profiling graph, $(\theta_1, \bft_{]1[}(\theta_1)=0)$ is a
straight line following this ridge, whereas the profiling graph
$(\bft_{]2[}(\theta_2),\theta_2)$ is showing a cusp.
The profiled posterior and marginal posterior distribution for
$\theta_2$ overlap, but for $\theta_1$ they show a qualitatively
different behaviour.
The Laplace approximation follows the marginal $p(\theta_2| \bfd)$
closely, for $\theta_1$ we observe a shift.
As already discussed the profile likelihood $\elL_p(\theta_1)$ is
divergent.  This is reflected in the behaviour of the profiled
posterior $p_p(\theta_1| \bfd)$ with its peak at the boundary.  The
marginalisation of the posterior distribution is able to regularise
this divergence and we obtain a bell-shaped marginal
posterior distribution $p(\theta_1| \bfd)$ with a mode
only slightly larger than the measured $x=1$.

Montoya et al.\,\cite{montoya:criticism} argue that example\,3 from
Berger et al.\,\cite{berger:integrated} (our
$\elL( \bfd |\theta_1,\theta_2)$ from Eqn.\,(\ref{eq:berger3})) is not
a valid likelihood function because it does not include the effect of
a finite precision of the measurements. We investigate this by adding
a constant $v$ to the variance of $y$, similar to the constant
component in $\sigma_2$ in Eqn.\,(\ref{eq:veff}).  Now for a finite
precision $v$ the criticism of Montoya
et~al.\,\cite{montoya:criticism} does not apply anymore.  Still the
profiled posterior distribution is approaching the distribution with a
peak at the boundary, for small $v$.  Similarly to the example in
section~\ref{sec:veff-model} the profiled posterior distribution is
strongly depending on the resolution $v$, whereas again the marginal
posterior distribution stays invariant.

Sometimes the volume effect is mentioned as reason why a profile
likelihood should be used instead of the marginal posterior
distribution. However, in this case only the marginalisation (the
"volume" integration) leads to a well-behaved distribution for
$\theta_1$, whereas the maximisation results in an almost singular
profiled posterior distribution.

\subsection{A Rosenbrock function}
\label{sec:banana}

\begin{figure}
  \begin{minipage}{0.4\textwidth}
    \includegraphics[width=\textwidth]{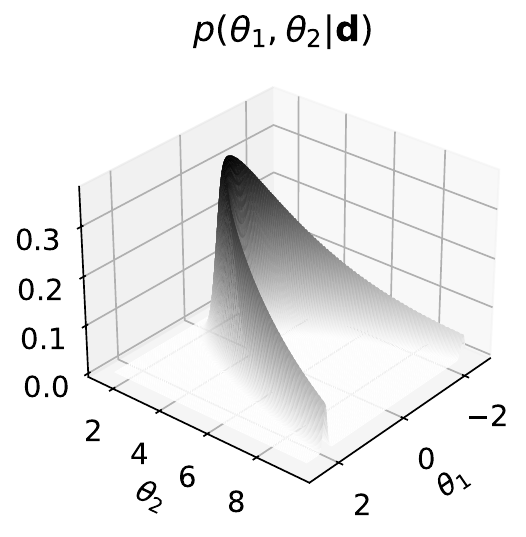}\\
    \includegraphics[width=\textwidth]{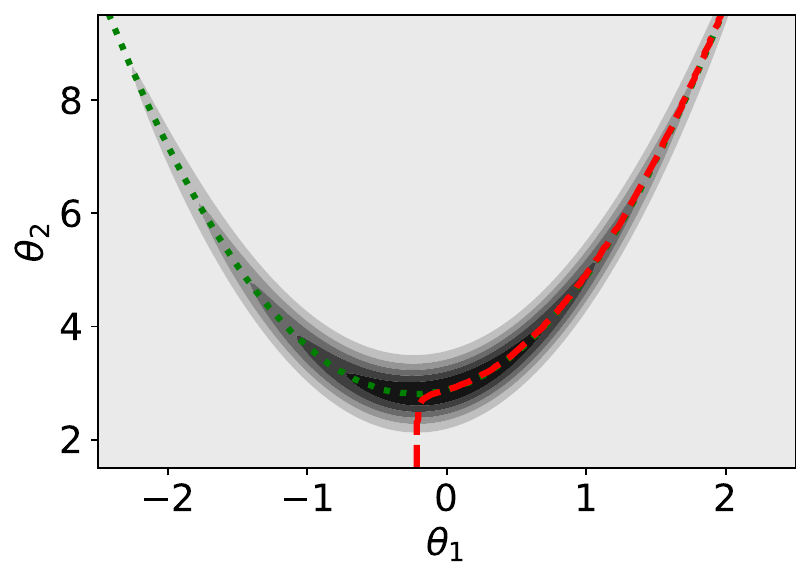}
  \end{minipage}
  \hfill
  \begin{minipage}{0.55\textwidth}
    \includegraphics[width=0.9\textwidth]{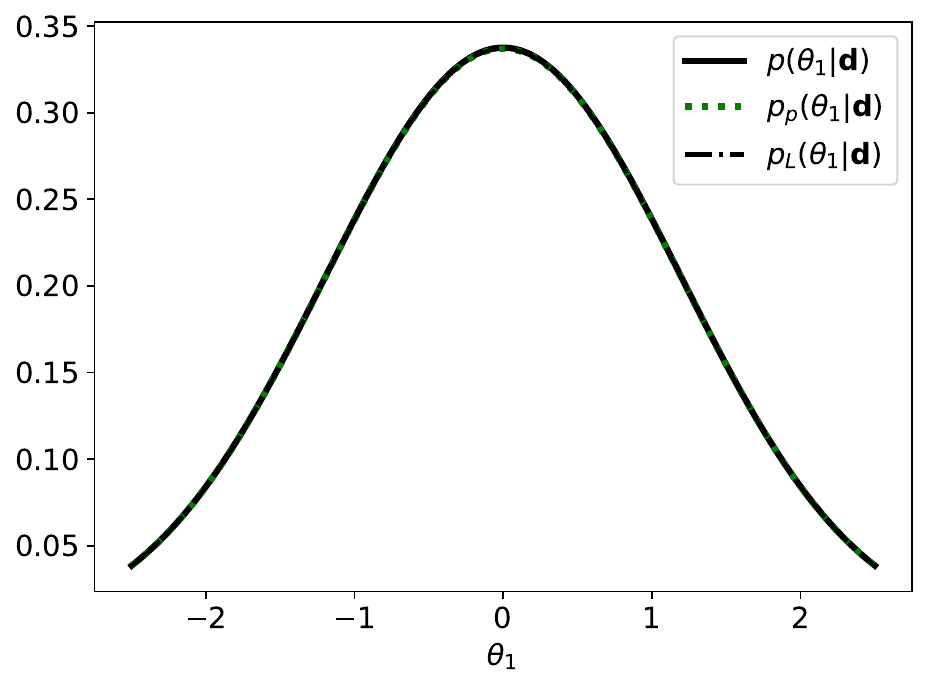}\\
    \includegraphics[width=0.9\textwidth]{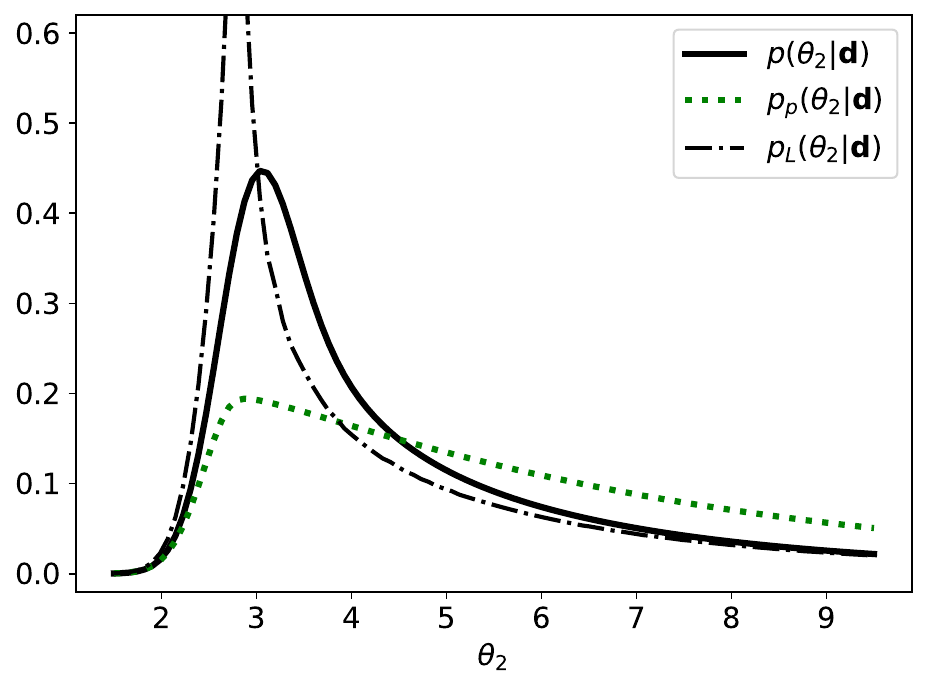}
  \end{minipage}
  \caption{On the left, the "banana-shaped" two-dimensional posterior
    $p(\theta_1,\theta_2|\bfd)$ is shown for the Rosenbrock
    example. In the contour plot, we also show the profiling graphs
    $(\theta_1, \bft_{]1[}(\theta_1))$ (green dotted) and
    $(\bft_{]2[}(\theta_2),\theta_2)$ (red dashed). On the right on
    top, we compare the marginal $p(\theta_1|\bfd)$ to the profiled
    posterior $p_p(\theta_1|\bfd)$. On the right, at the bottom, we
    compare the marginal $p(\theta_2|\bfd)$ with the profiled
    posterior $p_p(\theta_2|\bfd)$. The corresponding Laplace approximations
    $p_L(\theta_i|\bfd)$ are shown as dashed-dotted lines.}
  \label{fig:banana-example}
\end{figure}

Often the Rosenbrock function \cite{rosenbrock:automatic} is used to
illustrate convergence properties of Markov chain Monte Carlo
strategies (e.g.\ \cite{goodman:ensemble}). We use a similarly transformed
Gaussian as our parameterisation of such a "banana-shaped" distribution
\cite{wang:hamiltonian}.
The posterior $p(\theta_1,\theta_2| \bfd)$ is now proportional to
\begin{equation}
  p(\theta_1,\theta_2| \bfd) \propto
  \exp\left[ -\frac{1}{2(1-\rho^2)}
    \left(\frac{\theta_1^2}{a^2} + a^2 t^2 - 2\rho \theta_1 t \right)
  \right],
  \label{eq:banana}
\end{equation}
with $t=\theta_2 -\frac{b}{a^2} \theta_1^2 -b a^2$, the parameters
$a,b\in\mathbb{R}$, and the correlation coefficient $\rho$.
We calculate the marginal and the profiled posterior distributions as
described for the example in section\,\ref{sec:veff-model}.
In figure\,\ref{fig:banana-example} we show a slightly asymmetric
distribution $p(\theta_1,\theta_2| \bfd)$ obtained by using the
parameters $a=1.2$, $b=2$, and the correlation coefficient $\rho=0.9$.
For $\theta_1$ the marginal distribution $p(\theta_1|\bfd)$ shows a
bell shaped curve, matched by the profiled posterior
$p_p(\theta_1|\bfd)$.
For $\theta_2$ the marginal distribution $p(\theta_2|\bfd)$ is
 clearly skewed but the profiled posterior $p_p(\theta_2|\bfd)$
shows an even stronger skewing with a heavy tail for large $\theta_2$.
This is also the direction where the credibility region of the 
Rosenbrock function is curved inward (concave).  
The difference between the profiled posterior and marginal posterior
is not astonishing if one is looking at the profiling graph.  The
profiling graph $(\bft_{]2[}(\theta_2),\theta_2)$ is following only
one "branch" in the posterior distribution, whereas the marginal
posterior incorporates contributions from both branches.
The Laplace approximation is almost perfect for the marginal posterior
$p(\theta_1|\bfd)$.  However for $\theta_2$ the Laplace approximation
$p_L(\theta_2|\bfd)$ is missing the contribution from the second
branch of the posterior and is therefore shifted and significantly
tighter than the full marginal.
We checked with the hybrid Rosenbrock function of Pagani et
al.\,\cite{pagani:ndimensional} that also in higher dimensions
the profiled posteriors differ from the
marginal posterior distributions.

\subsubsection{A Rosenbrock distribution with added noise}
\label{sec:banana-noise}

\begin{figure}
  \begin{minipage}{0.4\textwidth}
    \includegraphics[width=\textwidth]{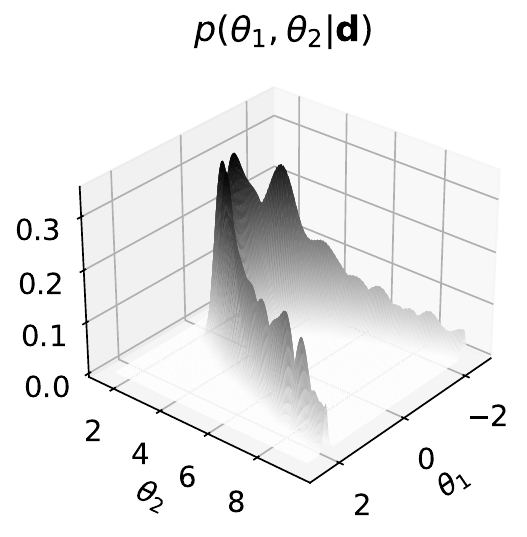}\\
    \includegraphics[width=\textwidth]{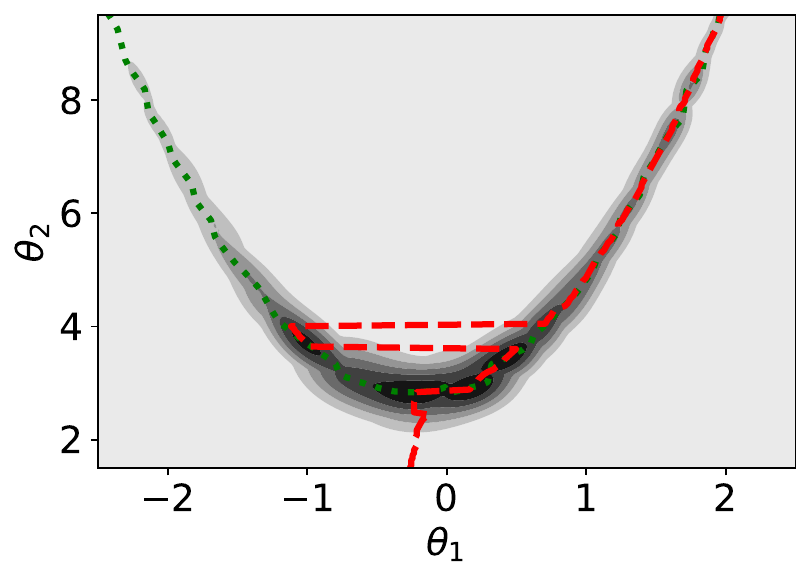}
  \end{minipage}
  \hfill
  \begin{minipage}{0.55\textwidth}
    \includegraphics[width=0.9\textwidth]{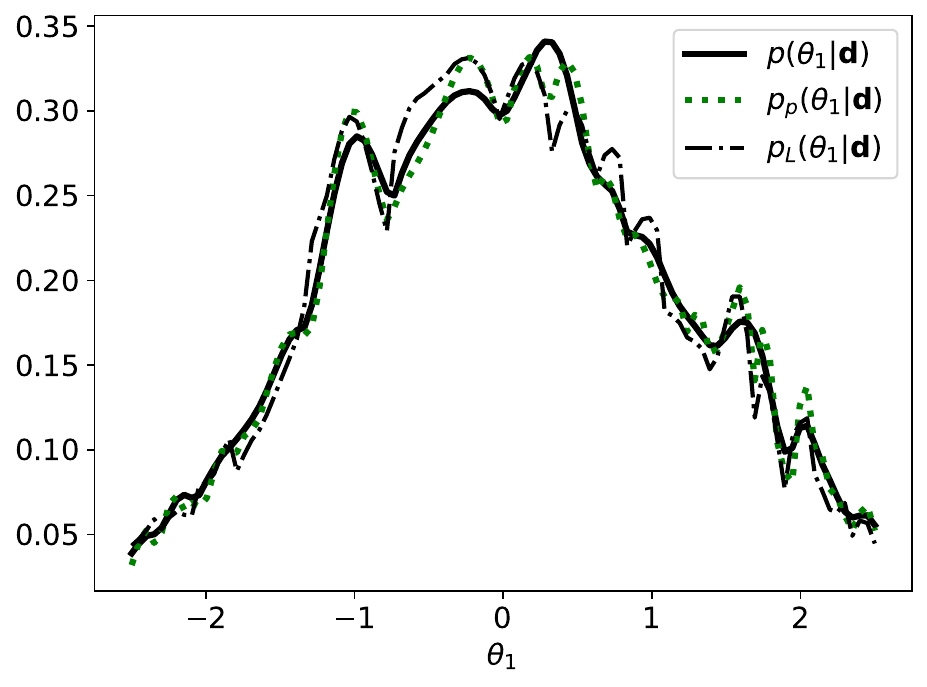}\\
    \includegraphics[width=0.9\textwidth]{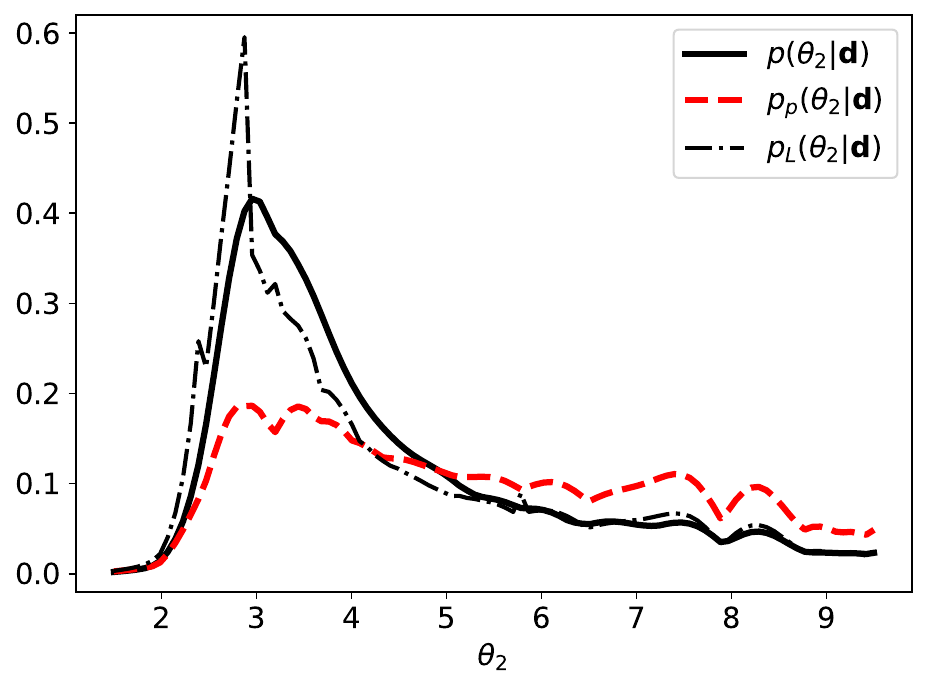}
  \end{minipage}
  \caption{Similar plots as in figure~\ref{fig:banana-example} but
    with added noise.}
  \label{fig:noisy_banana-example}
\end{figure}

With this Rosenbrock function, we can illustrate a further
characteristic trait of the profiling procedure.  Up to now, we
considered smooth posteriors in our examples. Since the likelihood 
and consequently the posterior distributions are depending on the
actual data they often exhibit a more bumpy shape. We emulate this by
adding noise to the posterior\,(\ref{eq:banana}) and re-normalise the
distribution.
Compare the posterior without noise in figure~\ref{fig:banana-example}
to the posterior with noise in figure~\ref{fig:noisy_banana-example}.
The overall characteristics of the one-dimensional marginal and
profiled posterior distributions do not change significantly by adding
the noise.  In addition, the profiling graph
$(\theta_1,\bft_{]1[}(\theta_1))$ for $\theta_1$ only shows small
wiggles in the noisy case.
But the profiling graph $(\bft_{]2[}(\theta_2),\theta_2)$ for $\theta_2$
is not continuous any more. It jumps from one branch to the other
and back (see figure~\ref{fig:noisy_banana-example}).
In the following analysis of the supernova data we observe a similar
behaviour.
The Laplace approximations show the same overall characteristics as
for the example without noise. Furthermore the jumps in the profiling
path bring about discontinuities in the Laplace approximations
$p_L(\theta_2|\bfd)$. We will see this effect even more pronounced in
the analysis of real data.

\section{Cosmological parameters from the SN\,Ia
  magnitude redshift  relation}
\label{sec:data}

The physical foundation of the following analysis is that type~Ia
supernova explosions essentially have the same peak
luminosity. Measurement of the magnitude and the redshift of such an
explosion gives information about the geometry of spacetime between
our position and the supernova.  This can be used to constrain
parameters of cosmological models for the spacetime.  For more
information on the method and on the history of this approach see the
review by S.\,Perlmutter~\cite{perlmutter:nobel}.
In the following we use the Pantheon+ supernova data set
\cite{scolnic:pantheon}.  Partly we redo the analysis of Brout et
al.\,\cite{brout:cosmology}. We confirm their results in the cases
where we investigate the same situations.  Additionally we focus on
the comparison between the marginal posterior and the profiled
posterior distribution.

\subsection{The sample, the models and the analysis}
\label{sec:detailsanalysis}

\begin{table}
  \caption{The cosmological models, their parameters and the 
    priors used for the MCMC calculations. ${\cal U}(a,b)$ is the
    flat\,/\,uniform distribution on the interval $[a,b]$.
    \label{table:models}
  }
\begin{center}
\begin{tabular}{|c|l|}
  \hline
   cosmology & priors \phantom{\LARGE I} \\
  \hline
  flat $w$CDM\phantom{\LARGE I}
    & $H_0\sim{\cal U}(10,200)$, $\Omega_m\sim{\cal U}(0.05,1)$,\quad 
      $\Omega_\Lambda=1-\Omega_m$,\quad $w\sim{\cal U}(-3,1)$  \\
  \hline
  non-flat $\Lambda$CDM\phantom{\LARGE I}
      & $H_0\sim{\cal U}(10,200)$, $\Omega_m\sim{\cal U}(0.05,1.5)$,
        $\Omega_\Lambda\sim{\cal U}(0,1.5)$, $w=-1$\\
  \hline
  non-flat $w$CDM\phantom{\LARGE I}
     & $H_0\sim{\cal U}(10,200)$, $\Omega_m\sim{\cal U}(0.05,1.5)$,
       $\Omega_\Lambda\sim{\cal U}(0,1.5)$, $w\sim{\cal U}(-3,1)$ \\
\hline
\end{tabular}
\end{center}
\end{table}

The Pantheon+ data set\footnote{The data is easily accessible at
  \url{https://github.com/PantheonPlusSH0ES/DataRelease}.}
\cite{scolnic:pantheon} includes the redshift $z_i$ and the distance
modulus $\mu_i$ of $N=1701$ type~Ia supernovae.  We use supernovae
with a redshift of $z>0.01$, hence we are left with 1590 data points.
To compare the corrected and standardised distance moduli from the
Pantheon+ sample with the theoretical predictions, we assume a
Gaussian likelihood where we use the provided covariance matrix
$\Sigma$ including systematic and statistical correlations. Then the
likelihood is
\begin{equation}
\label{eq:gauss-likelihood}
\elL(\bfd\,|\,\Theta) = 
\frac{1}{\left((2 \pi )^N \det(\Sigma)\right)^{\frac{1}{2}}} \ 
\exp\left[- \tfrac{1}{2}\
  \bfDelta_\mu(\Theta)^T \Sigma^{-1} \bfDelta_\mu(\Theta)\, \right] .
\end{equation}
$\mu_\text{model}(z,\Theta)$ is the predicted distance
modulus at redshift $z$ for a given model with parameters $\Theta$,
and the
$\bfDelta_\mu(\Theta)=\left( \mu_\text{model}(z_1,\Theta)-\mu_1,\ldots,
\mu_\text{model}(z_N,\Theta)-\mu_N \right)^T$.
Assuming a Friedman-Lemaitre-Robertson-Walker metric we can calculate
a theoretical prediction for the distance moduli for given redshifts
in several different cosmological models. As parameters
$\Theta$ we consider the matter content $\Omega_m$, the Hubble
parameter $H_0$, and depending on the extension of the model, also the
dark energy contribution $\Omega_\Lambda$ and the equation of state
parameter $w$ for the dark energy component.  
To numerically determine the distance modulus we use the tools
provided in Astropy \cite{astropy:2022}.
We fix the following parameters to the values determined in the
Planck\,18 analysis \citep{planck:planck18}: $T_\text{CMB}=2.7255 K$,
$N_\text{eff}=3.046$, $m_\nu=0.06\text{eV}$ (for one neutrino species,
the others are assumed mass-less). The mass parameter
$\Omega_m=\Omega_b+\Omega_\text{DM}$ includes the contribution from baryons
$\Omega_b=0.04897$ and dark matter $\Omega_\text{DM}$.
The properties of the cosmological models are summarised in
table~\ref{table:models}. 
We choose these flat priors since we want to make the connection with
the profile likelihood as described in section~\ref{sec:marginalised}.
For the flat $w$CDM model we have $\Omega_m+\Omega_\Lambda=1$ and for
the non-flat $\Lambda$CDM model we have a fixed $w=-1$. In the non
flat $w$CDM and $\Lambda$CDM models $\Omega_m$ and $\Omega_\Lambda$
may vary independently.

The posterior distributions of $H_0$, $\Omega_m$, $\Omega_\Lambda$, and
$w$ are estimated using Monte Carlo Markov chain calculations.  These
Markov chains are generated with the affine invariant ensemble sampler
emcee \cite{foreman-mackey:emcee}.
We use 10~parallel chains with at least 1\,Mio. up to 20\,Mio. of steps.
After visual inspection with a trace-plot we discard the burnin
phase. We made sure that the effective sample size (see
e.g.\,\cite{vehtari:rank}) was always larger than~$10^4$.  For the
improved convergence criterion $\hat R$ \cite{vehtari:rank} we have
$\hat R-1\le0.01$ for any of the parameters and for all the models
considered. This is an indication for well converged chains.
The marginal distributions are estimated from these Markov chains
using a kernel density estimate.

To obtain the profiled posteriors we use the downhill simplex
minimiser provided in SciPy \cite{virtanen:scipy}. In the minimisation
we use boundaries given by the limits of the uniform priors as shown
in table\,\ref{table:models}. To verify that we locate the global
maxima, we inspect the graphs visually and compare with the results
from a simulated annealing strategy. For the non-flat $w$CDM model we
use a brute force grid-search followed by a downhill simplex
algorithm.
Especially for a non-flat $w$CDM model we observe parameter combinations
without big bang singularity \cite{boerner:wasthere}. In such
situations the integration used to determine the distance moduli is
not converging.  We use this as a criterion to exclude these parameter
combinations and set the corresponding likelihood to zero (see also
\cite{overduin:evolution}).

\subsection{Marginal and profiled posterior 
  distribution of the parameters}

\begin{figure}
  \begin{center}
    \includegraphics[width=0.7\textwidth]{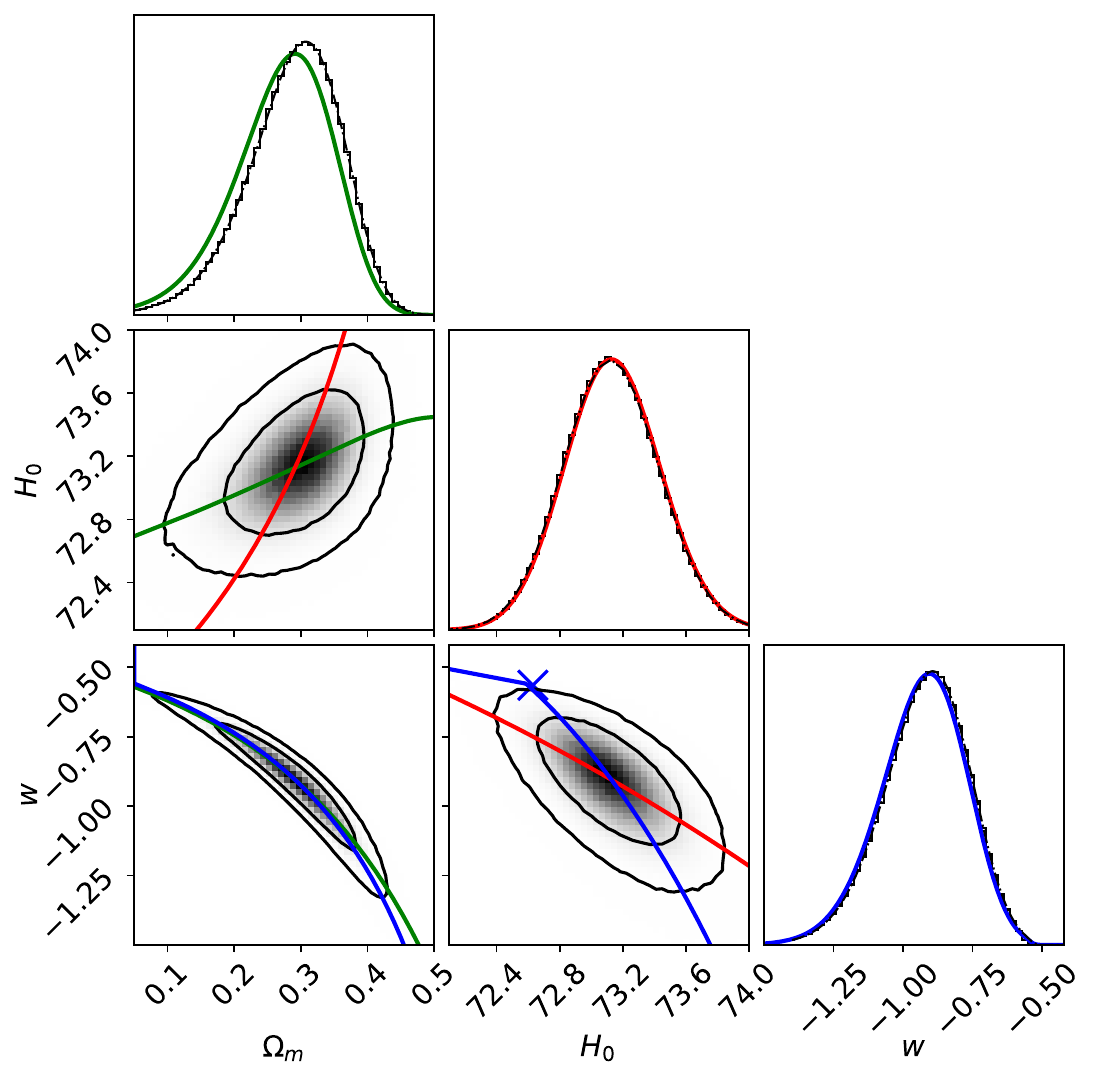}
  \end{center}
  \caption{The distribution of the cosmological parameters for the
    flat $w$CDM model. Marginal distributions are solid black. The
    profiled posterior distributions and the profile graph for
    $\Omega_m$ are in green, for $H_0$ in red, and for $w_0$ in blue. 
    The Laplace approximations are overlapping with the exact marginal
    posterior distributions.
    The kink in the profiling graph, due to the border
    $\Omega_m\ge0.05$, is marked with an $\times$. }
  \label{fig:modelI-I-corner}
\end{figure}

Figures~\ref{fig:modelI-I-corner}, \ref{fig:modelII-corner},
and~\ref{fig:modeIII-corner} show the one- and two-dimensional
marginal posterior distributions of the cosmological parameters
obtained for our models, together with the one-dimensional profiled
posterior distributions and the profiling graphs. The two contour
levels shown in the figures are the highest density credible
regions that include 68\% and 95\% of the posterior mass.
From the marginal posterior distribution, we confirm the estimates of
the cosmological parameters as given in table~3 of Brout et
al.~\cite{brout:cosmology} for the flat $\Lambda$CDM (not shown here),
the flat $w$CDM and the non-flat $\Lambda$CDM models (compare also
with figure~9 from Brout et al.~\cite{brout:cosmology}).  A discussion
of these results and their implications for cosmology are given in
Brout et al.~\cite{brout:cosmology}.
Here we compare the marginal with the profiled posterior
distributions.

\subsubsection{Flat\,$w$CDM and non-flat\,$\Lambda$CDM}

\begin{figure}
  \begin{center}
    \includegraphics[width=0.7\textwidth]{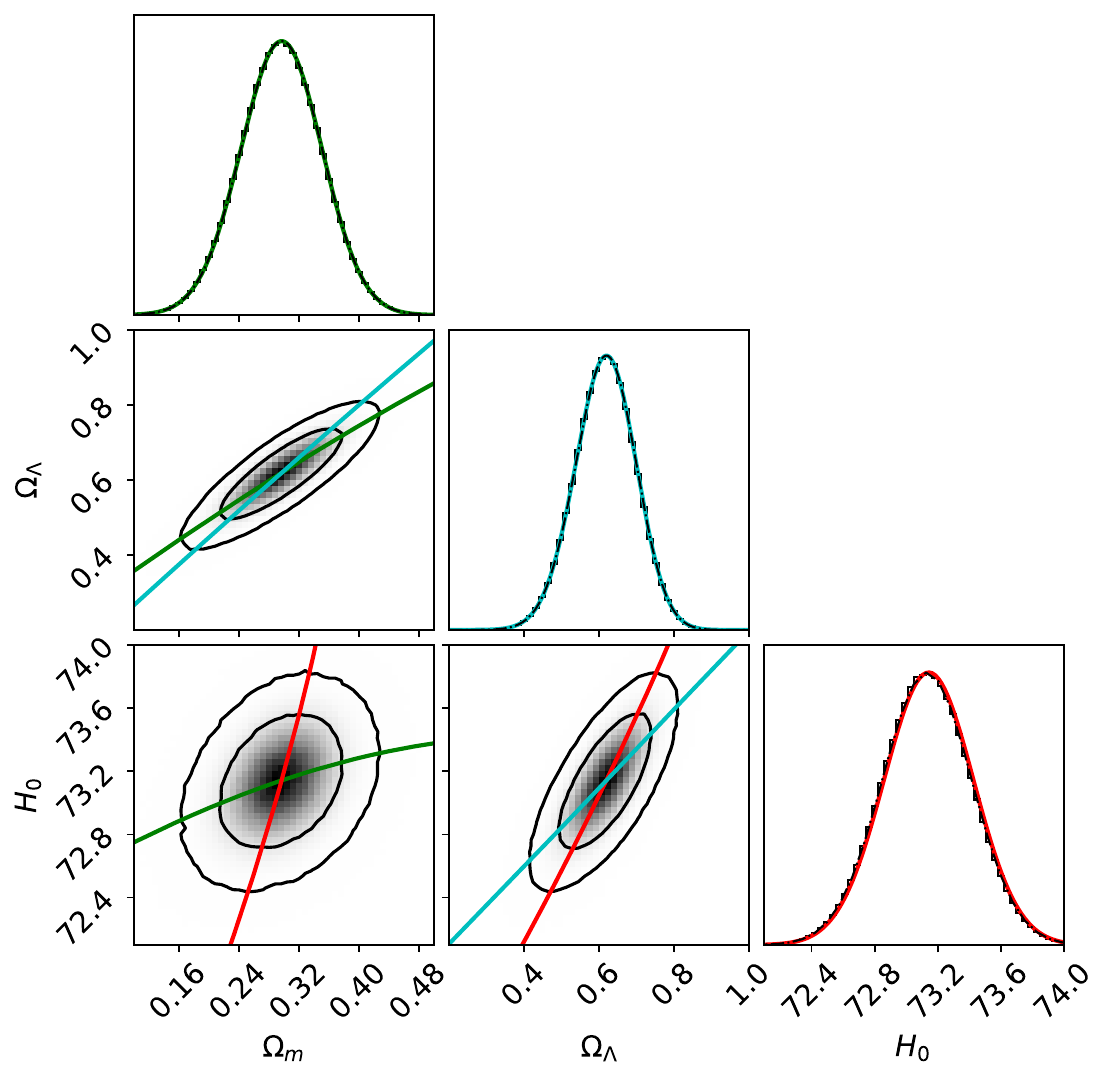}
  \end{center}
  \caption{The distribution of the cosmological parameters for the
    non-flat $\Lambda$CDM model. Marginal distributions are solid
    black.  The profiled posterior distribution and the profiling
    graph for $\Omega_m$ is in green, for $H_0$ in red, and for
    $\Omega_\Lambda$ in cyan.
    The Laplace approximations are overlapping with the exact marginal
    posterior distributions.}
  \label{fig:modelII-corner}
\end{figure}

As can be seen in the figures~\ref{fig:modelI-I-corner} 
and~\ref{fig:modelII-corner} the profiled posterior and
the marginal posterior distributions
are almost identical for the flat $w$CDM model and the 
non-flat $\Lambda$CDM model. Only in the $w$CDM model we observe 
a shift towards lower
values for $\Omega_m$ in the profiled posterior distribution
$p_p(\Omega_m|\bfd)$ in comparison to the marginal posterior
distribution $p(\Omega_m|\bfd)$.
As expected from the example in section\,\ref{sec:banana} we observe a
non-convex credible region in the two-dimensional marginal posterior
for $\Omega_m$ and $w$.  For the other parameters, the profiled and
marginal distributions overlap, and also the credible regions are
convex and almost elliptical.
%
%
For the flat $w$CDM model and the non-flat $\Lambda$CDM model 
the Laplace approximation to the marginal posterior distribution is perfect.
Nearly all the profiling graphs for the flat $w$CDM model
(figure\,\ref{fig:modelI-I-corner}) and the non-flat $\Lambda$CDM model
(figure\,\ref{fig:modelII-corner}) are smooth curves. Only for the flat
$w$CDM model we observe a kink in a profiling graph.  At this point,
the minimiser, used in the profiling procedure, hits the boundary
enforcing $\Omega_m\ge0.05$.  Because $\Omega_m$ also includes the
baryonic component we do not allow smaller values for $\Omega_m$.

\subsubsection{Non-flat\,$w$CDM}
\label{sec:nonflat-wCDM}

\begin{figure}
  \begin{center}
    \includegraphics[width=0.9\textwidth]{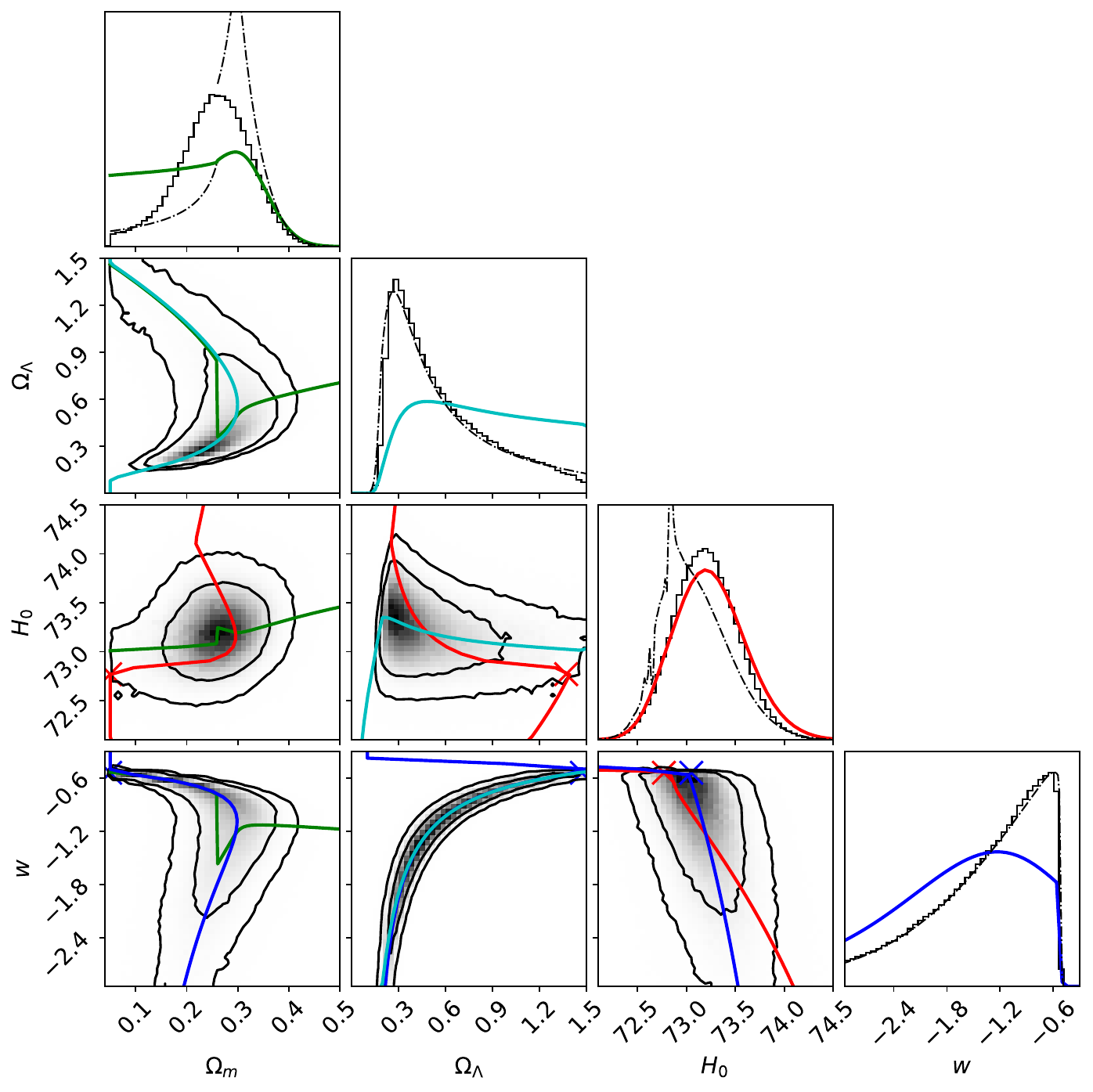}
  \end{center}
  \caption{ The distribution of the cosmological parameters for the
    non-flat $w$CDM model. Marginal distributions are solid black.
    The profiled posterior distribution and the profiling graph for
    $\Omega_m$ are in green, for $H_0$ in red, for $w_0$ in blue, and
    for $\Omega_\Lambda$ in cyan.  The kinks in the profiling graphs,
    due to the border $\Omega_m\ge0.05$, are marked with
    $\times$-es.
    The Laplace approximations for the marginal distributions are
    shown as dashed-dotted lines. The jump in $p_L(\Omega_m|\bfd)$ is
    \emph{not} a numerical artefact.}
  \label{fig:modeIII-corner}
\end{figure}

The overall agreement between the profiled and marginal posterior
distribution of the cosmological parameters dwindles away when we look
at the results from a non-flat $w$CDM model in
figure~\ref{fig:modeIII-corner}. Here we are going beyond the analysis
presented in Brout et al.~\cite{brout:cosmology}.
However, the marginal posterior distribution of $H_0$ is hardly
affected by the introduction of additional parameters and is in
agreement with the profiled distribution.
The marginal posterior distribution of $\Omega_m$ is still bell shaped
but the marginal posterior distributions of $\Omega_\Lambda$ and $w$
are heavily skewed. The marginal posterior distribution of $w$ peaks
at a value greater than $-1$ but has a significant weight in the
region smaller than $-1$.  For $\Omega_\Lambda$ values smaller than
0.5 are also possible, but still $\Omega_\Lambda$ is clearly bounded
away from zero with almost no weight below a value of $0.2$.
The two-dimensional marginal posterior distributions of
$\Omega_m$--$w$, $\Omega_m$--$\Omega_\Lambda$ and
$\Omega_\Lambda$--$w$ show distinct non-convex credible regions. As
expected from the discussion in section\,\ref{sec:banana} the one
dimensional profiled posterior distributions for $\Omega_m$,
$\Omega_\Lambda$ and $w$ show an even stronger skewing than already
visible in the corresponding marginal posterior distributions.
Indeed, the profiled posterior for $\Omega_\Lambda$ is an almost flat
distribution for $0.4<\Omega_\Lambda$ up to and beyond 1.0, offering
no further constraints. In addition, $\Omega_m$ and $w$ are
significantly less constrained by the profiled posterior than by the
marginal posterior.
The profiling graphs in the $\Omega_m$--$\Omega_\Lambda$ and
$\Omega_m$--$w$ contour plots show jumps similar to the jumps in
the example from section\,\ref{sec:banana}.
Some of the profiling graphs also show kinks,
where the minimiser hits the boundary enforcing $\Omega_m>0.05$.
For this non-flat $w$CDM model the Laplace approximations
$p_L(\Omega_m|\bfd)$ and $p_L(H_0|\bfd)$ are shifted and narrower
compared to the exact marginal distributions. The jumps in the
profiling graph lead to discontinuities in these approximations.

In the data analysis of the Pantheon+ sample as well as in the
examples from section~\ref{sec:example} we always use flat priors to
be able to compare marginal posterior distribution with the profiled
posterior distribution on an equal footing.
In a Bayesian analysis one could use either prior distributions which
are motivated by subjective prior knowledge or objective priors,
constructed from an extremalisation or based on invariance
requirements. Then however the direct proportionality of the profiled
posterior distribution with the profile likelihood is lost.
Our aim  in this work is the comparison of profiling and
marginalisation and therefore we stick to the flat priors.

\subsection{Summary of the one-dimensional distributions}

\begin{figure}
  \begin{center}
    \includegraphics[height=0.32\textwidth]{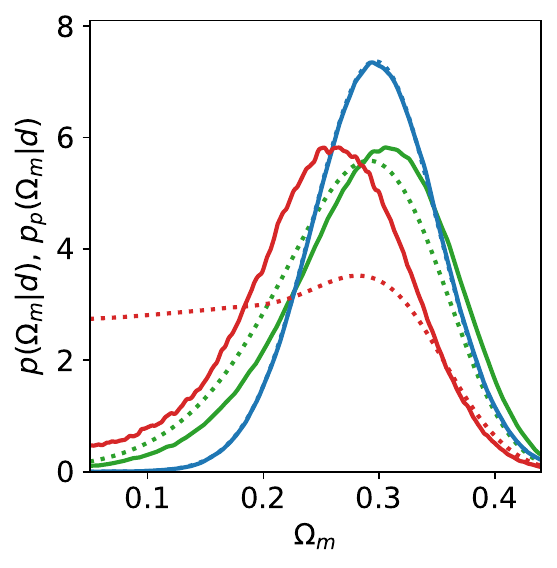}\hfill
    \includegraphics[height=0.32\textwidth]{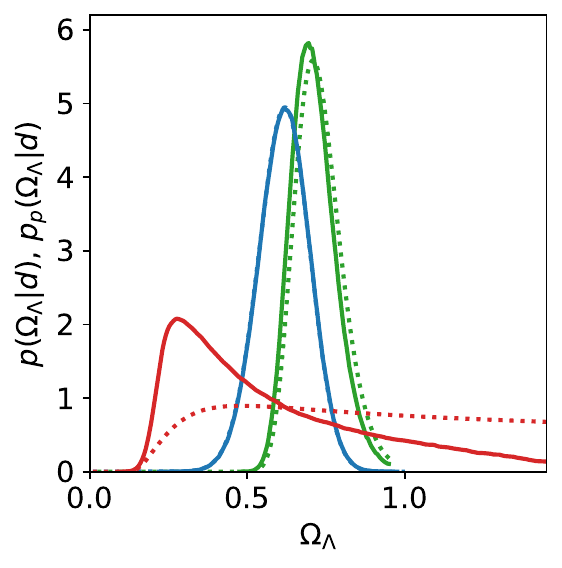}\hfill
    \includegraphics[height=0.32\textwidth]{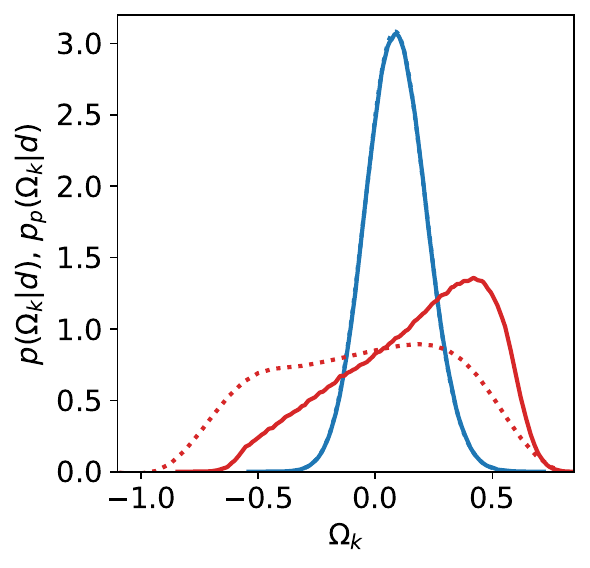}\\
    \begin{minipage}{0.19\textwidth}
    	\vspace*{-6cm}
    	\phantom{hallo}\includegraphics[width=\textwidth]{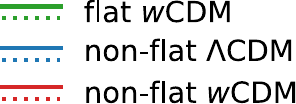}
    \end{minipage}
    \hspace{1.5cm}
    \includegraphics[height=0.32\textwidth]{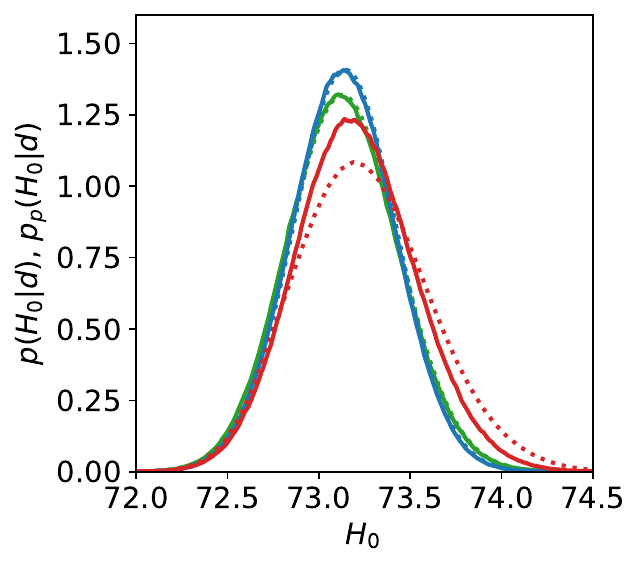}\hfill
    \includegraphics[height=0.32\textwidth]{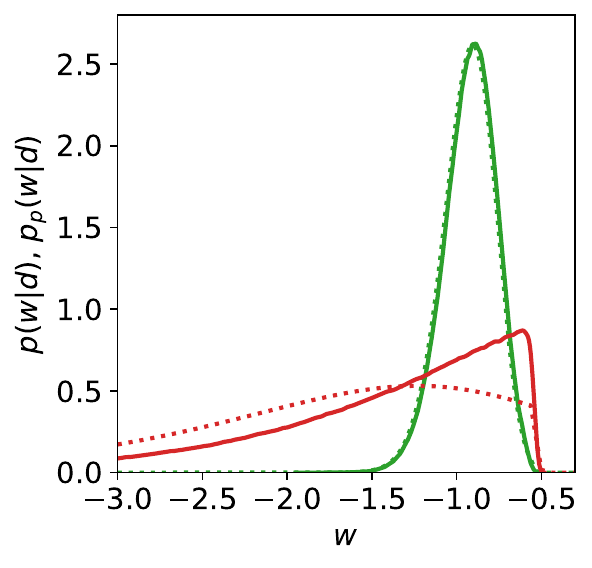}
  \end{center}
  \caption{Marginal posterior distribution (solid lines) and profiled
    posterior distributions (dotted lines) for the parameters in all
    our models.}
  \label{fig:densities}
\end{figure}

In figure\,\ref{fig:densities} we present a summary of the one
dimensional marginal distributions for $H_0$, $\Omega_m$,
$\Omega_\Lambda$, $\Omega_k$, and $w$ for all the models together with
the corresponding profiled posterior distributions.
The distribution of $H_0$ shows only small variation between the
models and between the marginal and profiled posterior distributions.
This is not surprising since the value of $H_0$ is determined mainly
by the local SHOES sample \cite{Riess:2.4}.
The distributions of $\Omega_m$ is constrained strongest in the
non-flat $\Lambda$CDM model.  In the flat $w$CDM the distributions are
consistent but slightly broader.  This is also the case for the
marginal posterior distribution in the non-flat $w$CDM model. But for
this model the profiled posterior distribution of $\Omega_m$ is
strongly skewed towards smaller values and we mainly get an upper
bound $\Omega_m\lesssim0.4$ from profiling.
The marginal posterior distribution for
$\Omega_k=1-\Omega_m-\Omega_\Lambda$ is readily obtained from the
Markov chains, but for the profiled posterior distribution we have to
start new profiling runs.
Both for $\Omega_\Lambda$ and $\Omega_k$ the distributions differ
between the models.  Specifically for the non-flat $w$CDM model the
marginal posterior distributions of $\Omega_\Lambda$ and $\Omega_k$
are strongly skewed.  Still they offer some localisation in parameter
space.  In this cases however the profiled posterior are essentially
flat distributions offering only the bounds:
$0.2\lesssim\Omega_\Lambda$, and $\Omega_m\lesssim0.5$, and
$-0.8\lesssim\Omega_k\lesssim0.7$.
A similar behaviour can be seen in the distribution of $w$. The marginal 
and especially the profiled posterior distribution only give 
the constraint $w\lesssim-0.5$.
Showing the distributions of $\Omega_k$ is redundant, but it
illustrates nicely that the supernova magnitude redshift relation is
only weekly constraining the curvature if one allows for $w\ne-1$ (see
e.g.~\cite{kowalski:improved}).
Throughout we observe that the marginal posterior distributions
give tighter constraints than the corresponding profiled posterior
distributions.

The main aim of our article is to highlight the differences between
profiling and marginalisation.  Specifically the non-flat $w$CDM model
shows several of the problems with the profiling procedure.
From a physical perspective considering non-flat cosmological models
with a uniform $\Omega_k$ is only a first step. The curvature is a
dynamical quantity.  Using the scaling solution
\cite{buchert:correspondence} from the backreacktion approach
Desgrange et al.~\cite{desgrange:dynamical} fit the supernova data
without a dark energy component. Closely related, the timescape
cosmologies provide a similar fit
\cite{dam:apparent,lane:cosmological}.
Modelling the dark energy component with a constant equation of state
parameter $w$ can be viewed as the simplest case of an effective
parameterisation \cite{chevallier:accelerating,linder:exploring}.
Further a parameterisations with orthonormal basis functions is
possible \cite{wagner:generalized}.
Certainly a physical model for the dark energy component is needed.
See Amendola et al.~\cite{amendola:euclid} for a review.

\section{Conclusions}
\label{sec:summary}

Instead of further summarising we will argue why one should use the
marginal posterior distributions instead of the profiled posterior
distribution or the profile likelihood to report results for parameter
estimates.

Let us start with a formal argumentation based on the results from
section\,\ref{sec:marginalised}. With the profiled posterior
distribution we construct a Bayesian analogue of the profile
likelihood. We show that the profiled posterior is a marginal
posterior distribution of a Bayesian hierarchical model.  The prior
distribution of this hierarchical model forces the parameters of the
model onto the profiling graph.
We use flat prior distributions in all the calculations of the
marginal posterior distribution presented here, since we want to
compare with the profiled
posterior distribution on an equal footing. 
Disregarding this connection we certainly may use different priors.
The prior used for the hierarchical model from
sect.\,\ref{sec:hierarchical} is one possibility. For this special
prior the marginal posterior distribution equals the profiled
posterior distribution.  If we want to make statements about the
parameters of the model by using the profiled posterior (respectively
the profile likelihood) we also should explain why we choose this
special prior.  It is not clear to us what physical effect we can
state, which is forcing the parameters onto the profiling graph.

Now let us consider the examples.  In section\,\ref{sec:example} we
identify two scenarios where the marginal differs from the profiled
posterior distribution.
First we construct a model with a tuneable volume effect. With this
nested model we are mimicking a situation observed for standard
$\Lambda$CDM model embedded into the early dark energy model.  We
can tune the accessible parameter space volume in the two dimensional
posterior distribution from a rather broad to a sharply peaked
distribution. Similarly we can tune the resolution (or the precision)
of the measurements in this example.
The marginal posterior distribution stays invariant under this tuning,
whereas the profiled posterior distribution is strongly depending on
the available parameter space volume or the resolution.
For a typically situation of parameter estimation the likelihood is
given by the experimental or observational setup. We have no tuneable
factor determining the available parameter space. The precision is
determined by the measurement or observation procedure.  Hence the
marginal posterior distribution seems to be the appropriate choice for
reporting results, if we want to be insensitive to variations of the
resolution or the parameter space volume.

We also discuss a closely related model of Berger et
al.\,\cite{berger:integrated}. Here the maximum of the posterior is on
a ridge-like structure which itself is \emph{not} surrounded by a
region where a larger fraction of the posterior mass is accumulating.
This maximum might not contribute appreciably to a marginal posterior
distribution.  However in this model the marginalisation, i.e.\ the
volume integration, leads to a well defined marginal posterior
distribution, whereas the maximisation results in an almost singular
profiled posterior distribution.

From the examples and the Pantheon+ data analysis in
section\,\ref{sec:data} we see that if the credible regions have a
simple convex shapes then the profiled posterior and the marginal
posterior distributions almost overlap.  As soon as we observe
non-convex credible regions, the profiled posterior and the marginal
posterior distributions start to deviate. In some cases they even show
a qualitatively different behaviour.
We demonstrate discontinuous jumps in the profiling graph, which are
not a numerical glitch, but an artefact of the method.
An overly conservative approach might be to accept results only if the
profiled and marginal posterior distributions agree.
For the Rosenbrock function in section~\ref{sec:banana} and in the
data analysis for the non-flat $w$CDM model in
section~\ref{sec:nonflat-wCDM} we have non-convex credible regions and
we observe skewed marginal posterior distributions. Still these
marginal posterior distributions offer some localisation in parameter
space.  However the corresponding profiled posterior distribution are
essentially flat and offer only bounds.
In addition, the ridge example in section~\ref{sec:ridge} shows that 
the profiling can lead to an almost diverging profiled posterior 
distributions.
Again we are led to the conclusion that the marginal posterior
distribution is the stable and preferable way to report results for a
parameter estimation.

\acknowledgments
It is a pleasure to thank Steffen Hagstotz for comments and suggestions.
This work would not have been possible without several libraries and
tools. We would like to thank the creators of these tools,
specifically we used Python with NumPy \citep{harris:numpy}, SciPy
\citep{virtanen:scipy}, numdifftools \cite{brodtkorb:numdifftools},
and Astropy\footnote{http://www.astropy.org} \citep{astropy:2013,
  astropy:2018,astropy:2022}.  We used the affine invariant ensemble
sampler emcee \cite{foreman-mackey:emcee} to generate the Markov
chains and ArviZ \citep{Kumar:arviz} to analyse them.  For plots we
were employing matplotlib \cite{Hunter:matplotlib} and
\texttt{corner.py} \cite{foreman-mackey:corner}.


\providecommand{\href}[2]{#2}\begingroup\raggedright\endgroup

\bibliographystyle{JHEP}
\bibliography{my}

\end{document}